\RequirePackage[hyphens]{url}
\documentclass[12pt]{article}
\usepackage[margin=1in]{geometry}
\usepackage{amsmath}
\usepackage{amsthm}
\usepackage{graphicx}
\usepackage[official]{eurosym}
\usepackage{url}
\usepackage{tablefootnote}
\usepackage{multirow}
\usepackage{subfig}
\usepackage{float}
\usepackage[hidelinks]{hyperref}
\usepackage{url}
\usepackage{breakurl}
\hypersetup{
    colorlinks=true,
    linkcolor=blue,
    filecolor=magenta,      
    urlcolor=cyan,
}
\usepackage{caption}
\usepackage{booktabs}
\usepackage{float}
\providecommand{\keywords}[1]{\textbf{keywords ---} #1}

\title{Analysis of ELSA COVID-19 Substudy response rate using machine learning algorithms}

\begin{document}

\author{Marjan Qazvini \footnote{Correspondence to: Email: marjan.qazvini@gmail.com, This paper is based on: \url{https://www.bayes.city.ac.uk/__data/assets/pdf_file/0007/700666/QAZVINI-Marjan.pdf}} (January, 2023)}

\date{}

\maketitle

\begin{abstract}
National Statistical Organisations every year spend time and money to collect information through surveys. Some of these surveys include follow-up studies, and usually, some participants due to factors such as death, immigration, change of employment, health, etc, do not participate in future surveys. In this study, we focus on the English Longitudinal Study of Ageing (ELSA) COVID-19 Substudy, which was carried out during the COVID-19 pandemic in two waves. In this substudy, some participants from wave 1 did not participate in wave 2. Our purpose is to predict non-responses using Machine Learning (ML) algorithms such as K-nearest neighbours (KNN), random forest (RF), AdaBoost, logistic regression, neural networks (NN), and support vector classifier (SVC). We find that RF outperforms other models in terms of balanced accuracy, KNN in terms of precision and test accuracy, and logistics regressions in terms of the area under the receiver operating characteristic curve (ROC), i.e. AUC.  
\end{abstract}


%\subsection*{Statement of significance}
%Maintaining the response rate is one of the challenges of follow-up studies. Understanding factors that affect the response rate and predicting the response rate can help with designing and budgeting surveys. In this study, we compare a variety of Machine Learning models and algorithms to predict non-response rates and use the permutation method to identify significant features.

\section{Introduction}
It is argued that Statistical Organisations that carry out follow-up surveys can reduce the cost by identifying the groups who may not participate in the future study. The identification of the factors that affect the non-response in a survey and the prediction of the non-response rate are problems that statistical organisations are dealing with. To address this problem UK Office for National Statistics (ONS) applies a logistic regression model to predict the propensity to respond (ONS 2009). In 2012, the US Bureau launched a Kaggle competition for the development of a predictive model of non-response. The winning models were based on ML ensemble models. Using random forest (RF) and gradient-boosted decision trees (GBDT), several factors were specified for the non-response. Later Erdman and Bates (2017) incorporate these factors into an ordinary least square regression (OLS) model and introduce a new metric, known as the low response score (LRS). They argue that, unlike ML models, traditional regression models are interpretable and hence more suitable for the industry. Bates and Mulry (2011) apply cluster analysis and introduce eight US geographic segments which were mutually exclusive. They show that the demographic, housing, and socioeconomic profile of these segments can determine the propensity to response in these geographic areas. Kulzick et al. (2019) apply models such as OLS, least absolute shrinkage and selection operator (LASSO), RF, and GBDT to predict and understand patterns of self-response in the 2020 Census where online and mail were two modes used by the census. They also select OLS despite lower predictive accuracy arguing that OLS is less complex and interpretable. They identify response mode, geography, and population demographics as factors that influence response rate. Zangger (2020) employs spatial models and through Monte Carlo simulation shows that these models can predict propensity to non-response well, provided that survey responses are spatially correlated. Ibrahim et al. (2021) apply non-parametric additive models with pairwise interactions to the US Census Bureau Planning database (PDB) and identify factors that affect the survey response rates. They compare their models with linear models such as LASSO and ridge regressions and ML models such as GBDT and NNs and conclude that the additive models not only are interpretable, but also their predictabilities are as good as the ML models. Mulry et al. (2021) examine the impact of lifestyle such as purchasing behaviour, leisure activities, etc on the propensity to non-response rate and find that these factors indeed affect the outcome. Australian Bureau of Statistics (2022) considers the problem of raising the response rate using RF with regression trees. They aim to predict the response rate by participants without a need for follow-up calls, i.e. gold providers as they call them.

\par In this study, we focus on the non-responses in the 2nd wave of ELSA COVID-19 Substudy which was carried out in 2 waves during the pandemic in 2020. The first wave was from June 3 to July 26 and the second wave was from November 4 to December 20. Since then, many studies investigate this dataset. In the following, we look at some of these studies. Di Gessa and Price (2021) use linear and logistic regression to study the effect of COVID-19 on health and social well-being of people with specific health problems and diseases such as diabetes, lung, and heart problems. They apply data from the first wave and the previous waves before the pandemic and find that those with health problems are more likely to experience poorer health and lower quality of life. Chen et al. (2022) investigate the impact of COVID-19 on factors such as difficulty with Activities of Daily Livings (ADLs), availability of ADL assistance, and inequality in the provision of ADL assistance using data from ELSA, the Survey of Health, Ageing and Retirement in Europe (SHARE) and the Health and Retirement Study (HRS) in the US. They apply country-specific weighted logistic regression models to study the factors that affect ADL assistance and Erreygers' corrected concentration index (ECI) to measure socio-economic inequality in receipt of ADL assistance. Some studies use this dataset to look at the mental health of the English population aged 50+ before and during the pandemic. See, e.g., Qin et al. (2022), Zaninotto et al. (2022), and Gaggero et al. (2022) among others. Wallinheimo and Evans (2021, 2022) investigate the relationship between the frequency and the purpose of using the internet and the quality of living and loneliness. They find that those who use the internet more frequently and for communication feel less lonely. See also, Iob et al. (2022), Curran et al. (2022), and the references therein.   

\par Research that analyses non-response in ELSA is scarce. For example, Cheshire et al. (2011) compare response rates between ELSA and the Health and Retirement Study (HRS). They find that the response rate in HRS is higher than ELSA due to factors such as sample design, respondent incentives, and interview mode. The role of incentives in increasing response rates has also been considered by Young et al. (2022) during the pandemic.  

\par Given the lockdown during the survey periods, the influence of factors that normally affect a survey response rate would be different. In particular, there are new factors that may affect the response rate such as COVID-19-related factors. Unlike previous ELSA waves, there is no end-of-life study to provide evidence regarding the causes of death. Our purpose is to predict the non-responses in the 2nd wave using ML algorithms and identify the factors that affect the non-responses in the 2nd wave. To the best of our knowledge, this is the first time that non-response is investigated in this dataset. The rest of this paper is organised as follows: Chapter 2 analyses this dataset. Chapter 3 describes the models and algorithms. Chapter 4 discusses the results and Chapter 5 concludes. All computations are implemented in Python - Scikit-learn.

\section{Data}
\label{sec:data}
ELSA is a self-reported survey conducted every two years and represents people aged 50 and over. The original sample was drawn from households that had previously participated in the Health Survey for England (HSE) between 1998 and 2001. Until now 9 waves have been carried out and new refreshments have been introduced in waves 3, 4, 6, 7, and 9. During the COVID-19 pandemic, a new substudy was designed, where participants were selected from the existing ELSA sample to investigate the socio-economic, mental health, physical health, COVID-19-related health, employment and demographics impacts of the lockdown and COVID-19 on people aged 50 and over. This survey was designed in modes of online and by phone. 

\par In this study, we consider ELSA \textit{core members}, i.e. participants who meet the age eligibility criteria of a given ELSA cohort, have participated in the HSE survey and the first wave of ELSA when invited to join the study. About $5,820$ and $5,594$ core members participated in wave 1 and wave 2, respectively, which is about $4\%$ drop in the number of responses in wave 2. We aim to investigate the factors that affect this drop in responses. This dataset is publicly available on the UKDS website. There are two files for each wave. We select core members from the variables \textit{FinStat} (wave 1) and \textit{Finstat\_w1} (wave 2). We then merge the two datasets and use the most frequent categories to fill in missing values for categorical features.

\begin{figure}[]
\centering
\includegraphics[scale = 0.47]{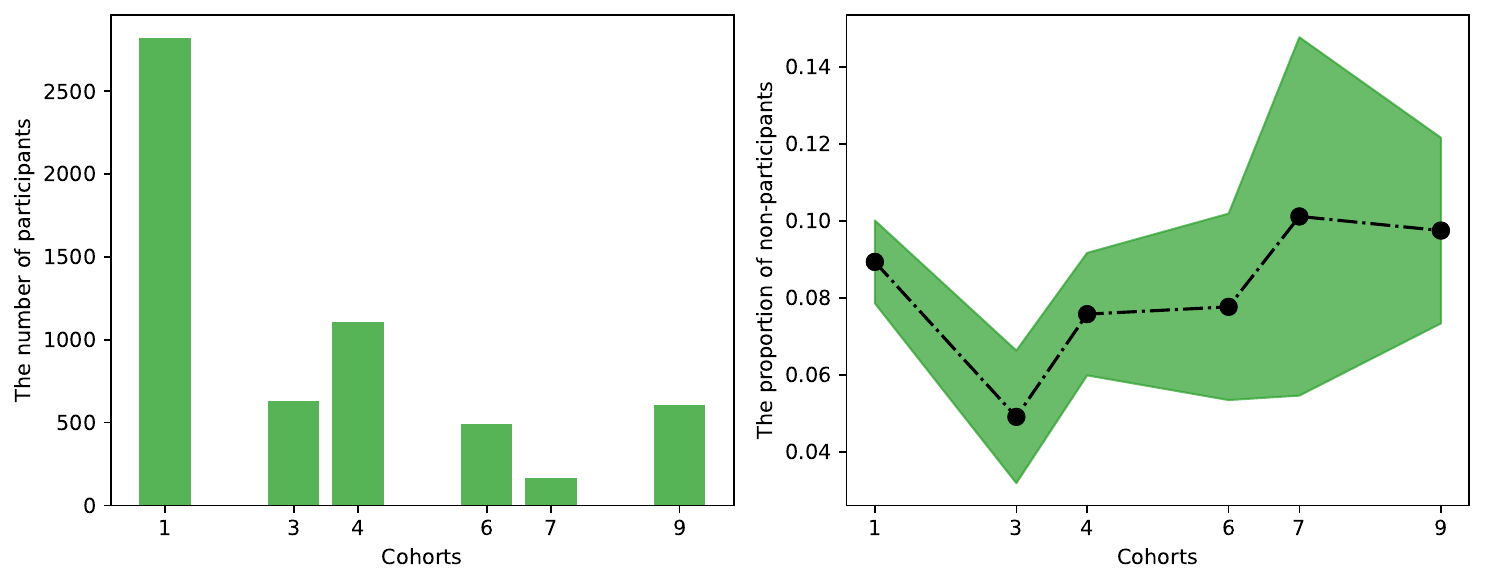}
\caption{The number of participants and proportion of non-participants per cohort}
\label{fig:cohort}
\end{figure}

\begin{figure}[]
\centering
\includegraphics[scale = 0.4]{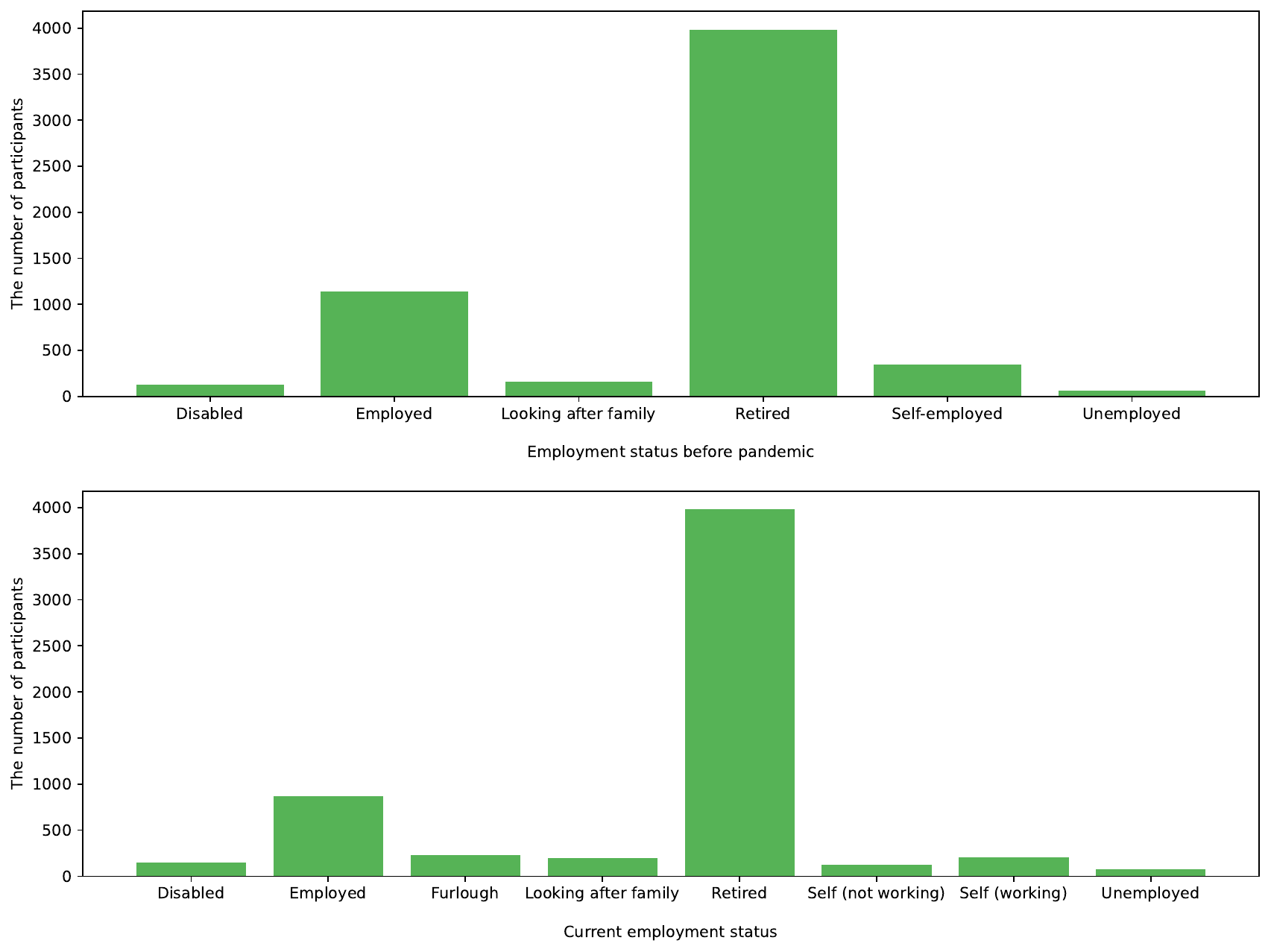}
\caption{Employment status before and during the pandemic among participants in wave 1 and non-participants in wave 2}
\label{fig:employment}
\end{figure}

\begin{figure}[]
\centering
\includegraphics[scale = 0.47]{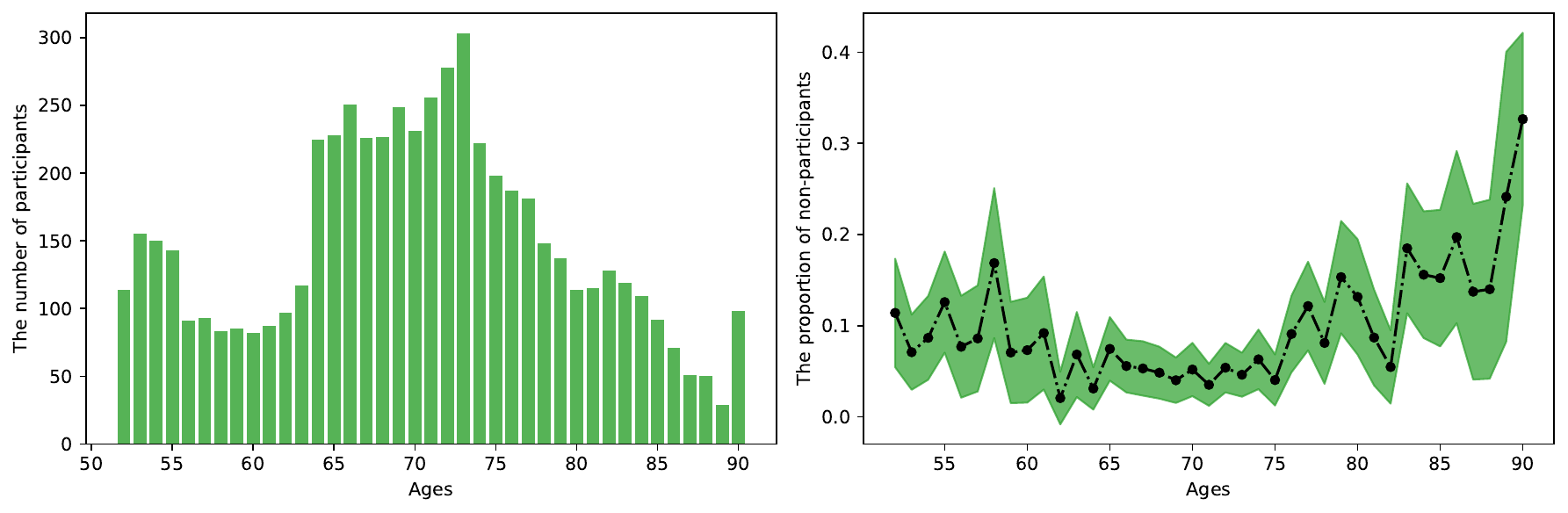}
\caption{The distribution of age among participants and proportion of non-participants at different ages}
\label{fig:age}
\end{figure}

\begin{figure}[]
\centering
\includegraphics[scale = 0.3]{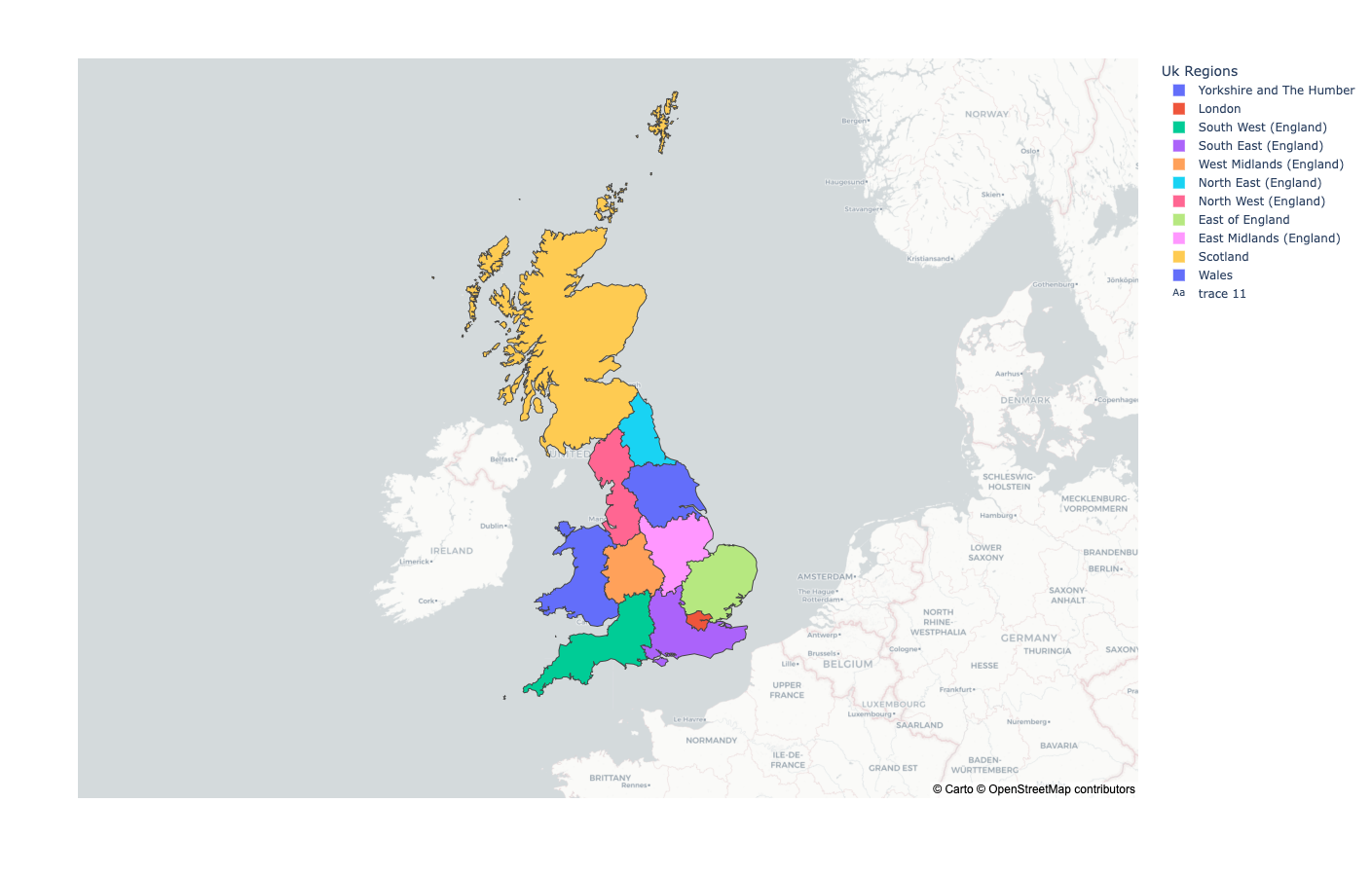}
\caption{Source: \url{https://geoportal.statistics.gov.uk}}
\label{fig:regions}
\end{figure}

\begin{figure}[]
\centering
\includegraphics[scale = 0.3]{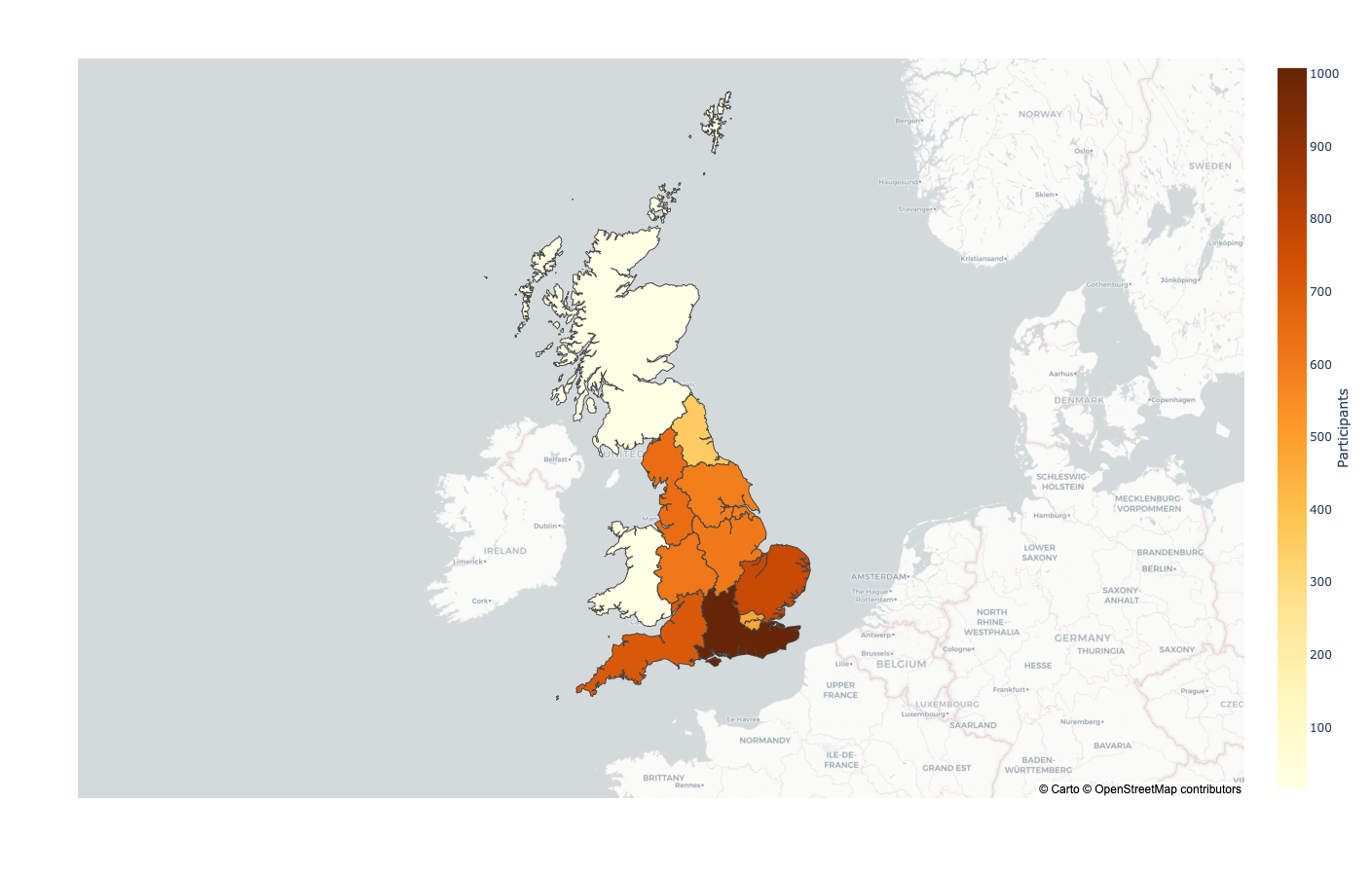}
\caption{The distribution of participants across UK regions}
\label{fig:participants}
\end{figure}

\begin{figure}[]
\centering
\includegraphics[scale = 0.3]{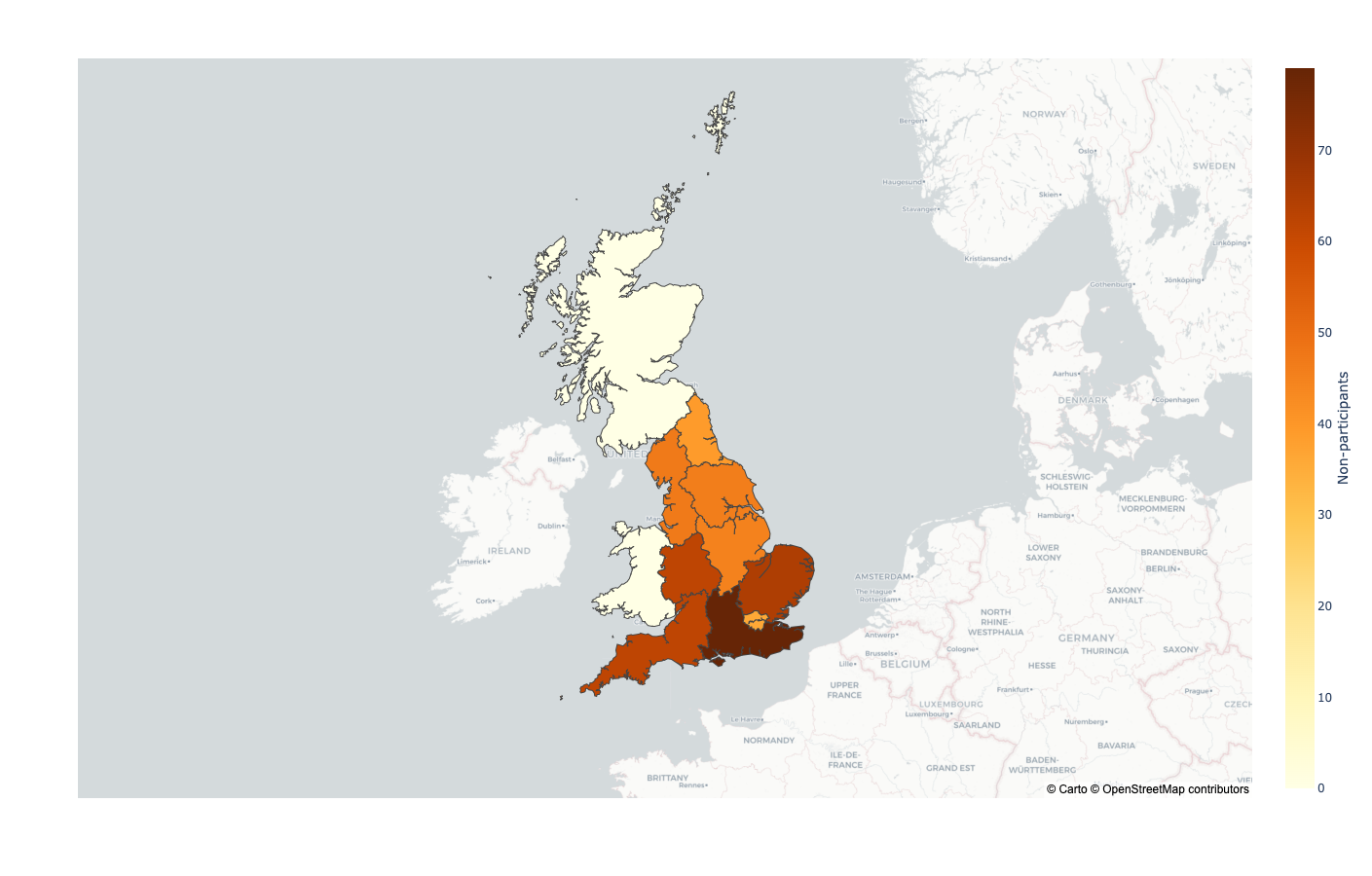}
\caption{The distribution of non-participants across UK regions}
\label{fig:nonparticipants}
\end{figure}

\begin{figure}[]
\centering
\includegraphics[scale = 0.3]{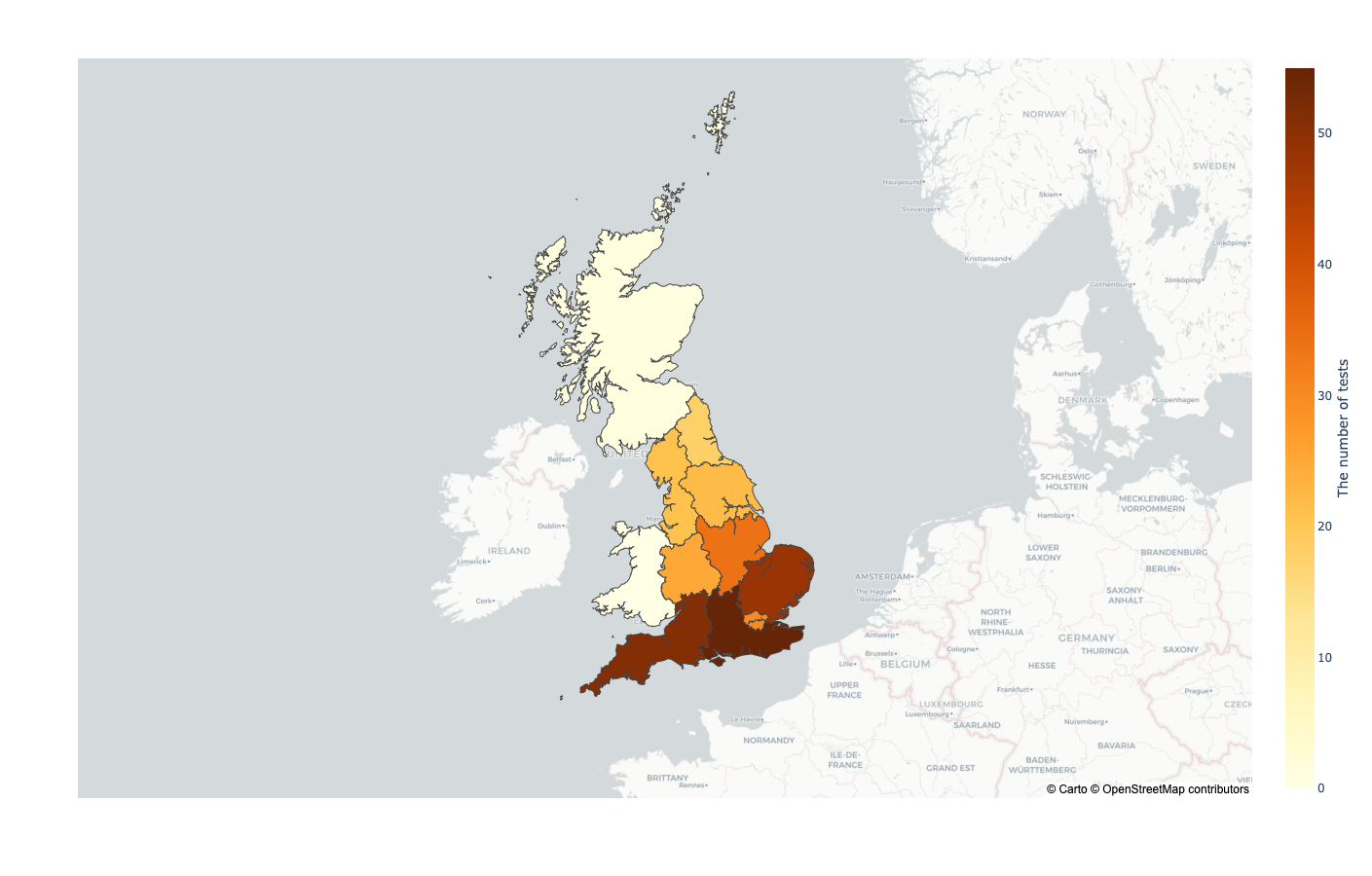}
\caption{The distribution of participants who were tested for COVID-19}
\label{fig:test}
\end{figure}

\begin{figure}[]
\centering
\includegraphics[scale = 0.3]{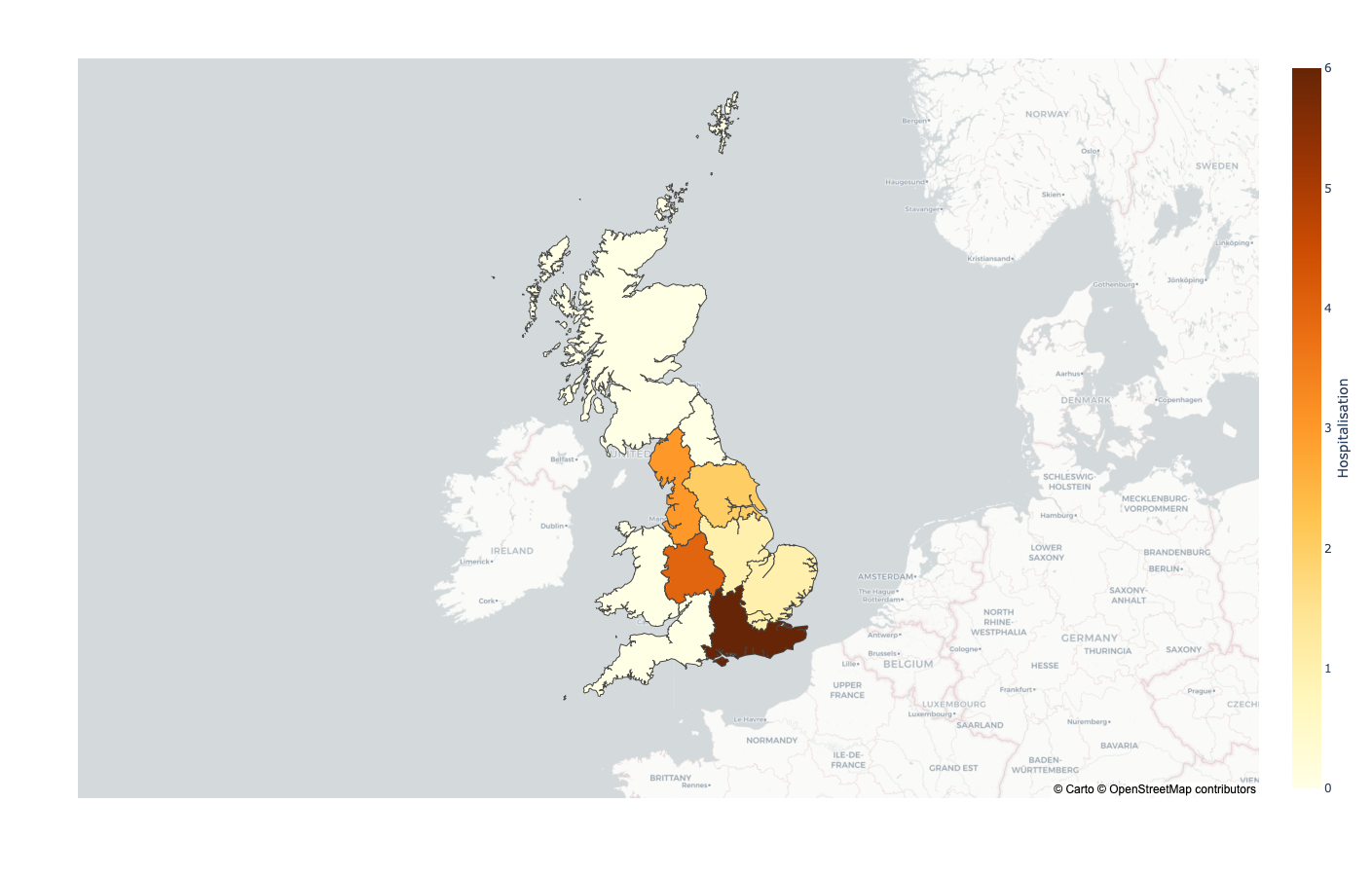}
\caption{The distribution of participants who were hospitalised due to COVID-19}
\label{fig:hospital}
\end{figure}

Table \ref{tab:cohorts} provides information about the participants from different cohorts. Most participants in wave 1 come from the first cohort. This is also illustrated in Figure \ref{fig:cohort}. In this figure, we obtain the confidence interval with the error $2 \sqrt{p(1-p)/n}$ where $p$ represents the proportion of non-participants per cohort and $n$, the number of participants per cohort. As we can see there are no particular trends between the proportion of non-participants and the cohorts that they come from. Figure \ref{tab:modes} shows the number of participants interviewed online and by phone per cohort. We can see that online is always preferred to phone. Another feature that we consider is the place of living during the pandemic. In Table \ref{tab:cohorts} we can see that most participants were living at their usual place during the pandemic and all individuals in a care home are from cohort 1. Most of the features that we consider in this study are dichotomous and related to COVID-19 symptoms and other health problems. In Section \ref{sec:results} we examine the impact of these health-related features on non-responses. During COVID-19 there were changes in the employment rate and economic activities 
\footnote{\url{https://www.ons.gov.uk/employmentandlabourmarket/peopleinwork/employmentandemployeetypes/bulletins/uklabourmarket/october2020} Accessed: December 2022}. Figure \ref{fig:employment} illustrates the employment status of participants before and during the pandemic. Due to the age group of participants, most of them are retired. We can observe that employment has dropped by $24\%$, those who were looking after their families have increased by about $35\%$, and $14\%$ more people become permanently sick or disabled. Table \ref{tab:cohorts} shows that about $10\%$ of participants in cohorts 9, 7, and 6 are smokers, whereas this is just $5\%$ in cohort 1. Figure \ref{fig:age} shows the distribution of age among participants. We can see that most participants are between ages 65-75 and as age increases, it is more likely that participants in wave 1 do not take part in wave 2. In Table \ref{tab:cohorts} we also see that most participants are from urban areas. The last feature that we consider is \textit{regions}. According to ONS \footnote{\url{https://www.ons.gov.uk/economy/environmentalaccounts/articles/leavingnoonebehindareviewofwhohasbeenmostaffectedbythecoronaviruspandemicintheuk/december2021} Accessed: December 2022} some groups have been more severely affected by COVID-19. Figure \ref{fig:regions} shows the map of the Nomenclature of Territorial Units for Statistics (NUTS) level 1 UK regions. We use this map to illustrate the distribution of participants and their responses across UK regions in this survey. Figure \ref{fig:participants} shows the distribution of participants in wave 1. As we can see most of the participants are from South East (England) and a small number of participants are from Scotland and Wales. We can observe a similar pattern in the distribution of non-responses in wave 2 in Figure \ref{fig:nonparticipants}. Figure \ref{fig:test} shows the distribution of participants who were tested for COVID-19 and Figure \ref{fig:hospital} shows the distribution of participants who were hospitalised due to COVID-19. In Section \ref{sec:results} we examine the significance of these features on non-response using the models and algorithms that we discuss in the next section.

%\begin{table}
%\caption{Interview mode in waves 1 and 2 and the number of core members}
%\centering
%\footnotesize
%\begin{tabular}{cccc}\toprule
%		& Online 		& Phone		& Core members \\\midrule
%Wave 1     & $5,791$		& $1,249$		& $5,820$ \\\midrule
%Wave 2	& $5,652$ 	& $1,142$		& $5,594$ \\\bottomrule
%\end{tabular}
%\label{tab:modes}
%\end{table}

\begin{table}
\caption{The number of responses from different cohorts}
\centering
\scriptsize
\begin{tabular}{lccccccc}\toprule
Cohorts							& 1    	&3		&4	 	&6 	       &7 		&9		& Total \\
								&2002-2003&2006-2007&2008-2009&2012-2013&2014-2015&2018-2019&\\\midrule
Number of responses in Wave 1     		& $2,819$ & $631$	& $1,108$ & $489$ &  $168$	&  $605$	& $5,820$\\
Number of non-responses in Wave 2		& $252$	& $31$	& $84$	& $38$	& $17$	& $59$	&  $481$\\\midrule
Interview mode: Phone				& $739$	& $68$	& $209$	& $47$	& $11$	& $61$	& $1,135$\\
Interview mode: Web				& $2,080$ & $563$	& $899$	& $442$	& $157$	& $544$	& $4,685$\\\midrule
Living at home						& $2,774$ & $621$	& $1,090$ & $482$	& $164$	& $598$	& $5,729$\\
Living in a care home				& $6$	& $0$	& $0$	& $0$	& $0$	& $0$	& $6$	\\
Living in hospital					& $1$	& $0$	& $0$	& $1$	& $0$	& $0$	& $2$	\\
In someone else's home				& $29$	& $6$	& $10$	& $4$	& $2$	& $3$	& $54$	\\
Somewhere else					& $9$	& $4$	& $8$	& $2$	& $2$	& $4$	& $29$	\\\midrule
Smokers							& $139$	& $51$	& $79$	& $51$	& $17$	& $64$	& $401$	\\
Non-smokers						& $2,680$ & $580$	& $1,029$ & $438$	& $151$	& $541$	& $5,419$	\\\midrule
Rural							& $780$	& $189$	& $302$	& $134$	& $41$	& $128$	& $1,574$\\
Urban							& $2,039$ & $442$	& $806$	& $355$	& $127$	& $477$	& $4,246$\\\bottomrule		
\end{tabular}
\label{tab:cohorts}
\end{table}

\begin{figure}[]
\centering
\includegraphics[scale = 0.47]{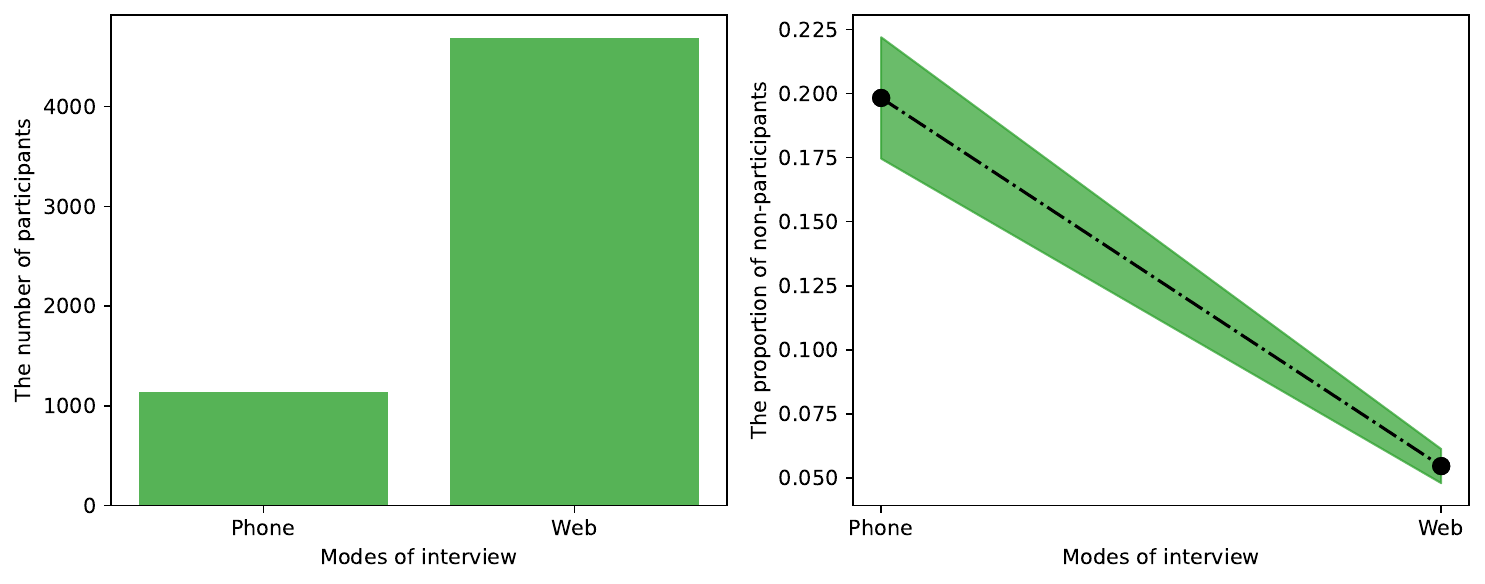}
\caption{The number of participants and proportion of non-participants by interview mode}
\label{fig:interview}
\end{figure}

\begin{figure}[]
\centering
\includegraphics[scale = 0.47]{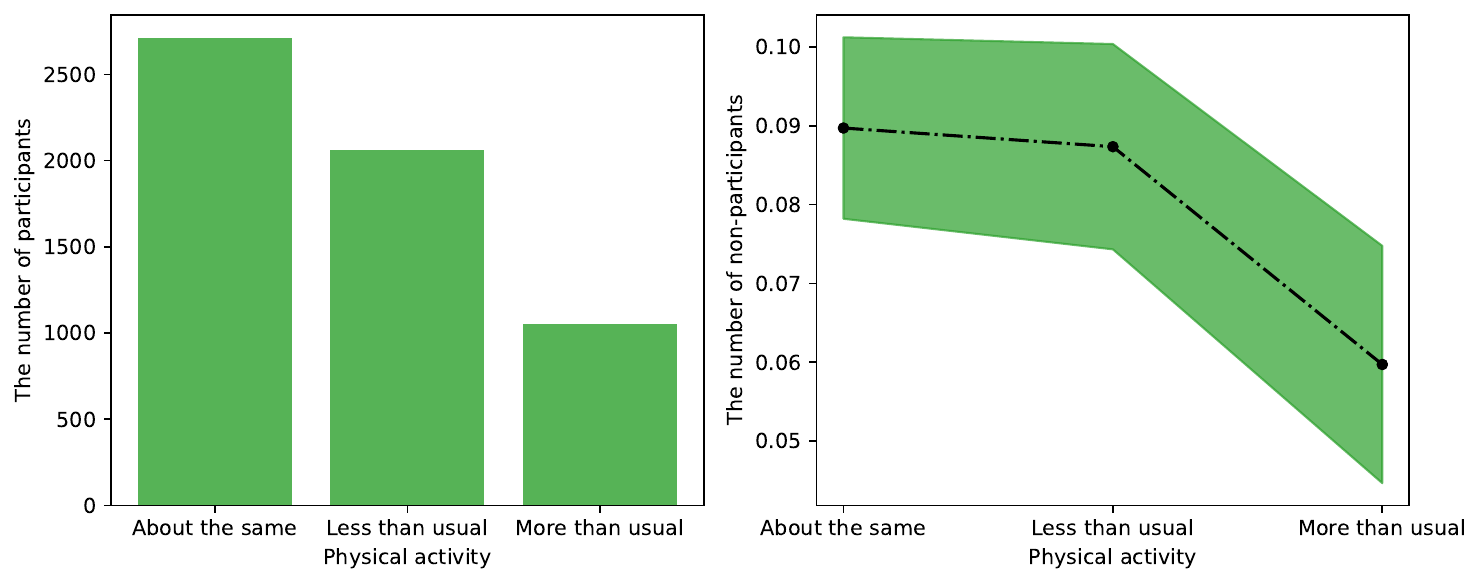}
\caption{Activity levels during the pandemic among participants in wave 1 and non-participants in wave 2}
\label{fig:activity}
\end{figure}

\section{Models and algorithms}
\label{sec:models}
Suppose $D = \{(x_i, y_i)\}_{i=1}^N$ is our training set, where $x_i \in R^d$ represent our participants with $d$ features (covariates), i.e. $x_i = (x_{i1}, x_{i2}, \dots, x_{id})$, $y_i \in \{0, 1\}$ are labels (class, dependent variables) and $N$ is the number of training examples. Our problem is a binary classification problem and we aim to learn a function (predictor, learner), $h$, that maps $x_i$  to $y_i$ where $y_i = 1$ represents \textit{non-response} and $y_i=0$ otherwise. In ML terms, $h$ is called a \textit{hypothesis}. Further, we assume that the training set is drawn from a joint probability distribution $P(\mathbf{x}, y)$. We define $L(y, h(\mathbf{x}))$ to be a non-negative real-valued loss function, which measures how the prediction of our hypothesis is different from the true value of $y$. In reality, the distribution of the training set is unknown and therefore we need to use an empirical distribution. Hence, we define the empirical risk (training error) by
\begin{eqnarray}
\label{eq:riskemp}
R(h) = \frac{1}{N} \sum_{i=1}^N L(y_i, h(\mathbf{x}_i)).\nonumber
\end{eqnarray}
Our goal is then to find a hypothesis among the hypothesis class $H = \{h_1, \dots, h_k\}$ that minimises this risk. After learning our predictor, we test it on an unseen dataset, called a test set. Training error is not typically close to the test error. If the training error is small and the test error is large, the model \textit{overfits} and has a large variance. Therefore, it does not generalise well to an unseen dataset. If the training error is large and the test error is also large, the model \textit{underfits} and has a large bias. One way to address the problem of over-fitting, as we see later, is regularisation, where we add a penalty term to the loss function. For more details, see, e.g., Hastie et al. (2009) and Murphy (2012). 

\subsection{K-nearest neighbours (KNNs) algorithm}
This algorithm was introduced by Fix and Hodges (1989) where they studied a discriminatory analysis in the case of two unknown distributions, namely, $F$ and $G$, and tried to determine which of these two distributions has generated an observed value $z$. KNNs algorithm is a non-parametric classifier, where the label of a data point (participant) in the test set is selected according to the labels of $K$ nearest points in the training set. Let $N_K(x, D)$ denote $K$-nearest points to $x$. Then in probabilistic form, we have
\begin{eqnarray}
\hat{y} = \arg \max_{y} p(y | \mathbf{x}, D)\nonumber,
\end{eqnarray}
where 
\begin{eqnarray}
p(y = 1 | \mathbf{x}, D) = \frac{1}{K} \sum_{i \in N_K(\mathbf{x}, D)} I(y_i = 1) \nonumber,
\end{eqnarray}
where $I$ is the indicator function, i.e. $I(A) = 1$ if $A$ is true. To find $K$ nearest points, we use a distance measure like Euclidean distance, which is a commonly used distance metric. 

\par In KNNs, $K$ is the hyperparameter that can be determined through cross-validation. The high value of $K$ leads to under-fitting and the small $K$ leads to over-fitting. Jiang et al. (2007) discuss the issues with KNNs and provide some suggestions on other distance measures and how to improve this algorithm.   

\subsection{Decision trees}
Morgan and Sonquist (1963) investigated the interaction effect among survey data by introducing the Automatic Interaction Detection (AID) algorithm, which is an algorithm for regression trees. Their algorithm was then extended by Messenger and Mandell (1972) to THeta Automatic Interaction Detection (THAID) which is an algorithm for classification trees. They looked at different measures and how to split a feature on. Later, two other algorithms, Induction Decision Tree (ID3) and C4.5 were introduced by Quinlan (1986) and Quinlan (1992), respectively. Quinlan used the ideas from information theory and split over features based on the maximum amount of information extracted given a particular feature. To avoid over-fitting, ID3 implements stopping criteria, whereas C4.5 suggests pruning i.e. deleting a portion of trees after growing a large tree. Classification And Regression Trees (CART) were introduced by Breiman et al. (1984). The idea of CART is based on AID and THAID. It applies a greedy algorithm, i.e. it looks for the best partition at each step and builds a full-grown tree, and then starts pruning (Loh, 2014).

\par In this study, we focus on CART (Breimen et al., 1984) methodology and in particular, binary classification trees. A decision tree is like an upside-down tree where we start with a root node and end with a terminal node or a leaf. CART is a greedy algorithm in which an optimal decision is chosen at each step. In CART, we partition the feature space into different regions denoted by $R$. Our goal is to minimise the misclassification rate or other criterion in each region by choosing the best feature to split on. Let $p_{m}$ denote the proportion of class $1$ in node $m$. We have
\begin{eqnarray}
p_{m} = \frac{1}{N_m} \sum_{x_i \in R_m} I (y_i = 1),  \quad N_m = \# \{i: x_i \in R_m\}\nonumber,
\end{eqnarray}
where $N_m$ is the number of observations in node $m$ and region $R_m$.

\par In order to find the best feature to split on, first we need a cost function and then we determine how a new feature leads to a reduction in cost, given by 
\begin{eqnarray}
\textnormal{cost}(D) - \left(\frac{N_L}{N} \textnormal{cost}(D_L) + \frac{N_R}{N} \textnormal{cost}(D_R)\right)\nonumber
\end{eqnarray}
where $N$, $N_L$, and $N_R$ are the number of observations in a root node, left node, and right node, respectively. Here, we use Gini index $2p (1-p)$. Decision trees are interpretable and good at understanding non-linear relationships. However, they are prone to over-fitting. To stop growing a tree we can pre-determine the number of points in $R$, maximum depth, etc. It is argued that combining different learners can improve their performance. We discuss this in the next section.

\subsection{Ensemble methods}
\label{sec:ensemble}
One way to reduce the over-fitting and hence the variance of an estimated predictor is to train multiple trees on subsets of training data chosen randomly with replacement, i.e. bootstrapping, and then select a class based on the majority vote. This method is known as bootstrap aggregating (bagging) and is introduced by Breiman (1996). One problem with bagging is that training trees based on bootstrap replicates leads to a high correlation among the trees. To address this issue Breiman (2001) proposes a method where trees are trained based on both a random sample of features and a random sample of training sets with replacement. This method which consists of unpruned trees is called random forest (RF). Another ensemble method introduced by Freund and Schapire (1997) is AdaBoost. In this method, multiple weak learners, i.e. those classifiers that only predict slightly better than chance, are combined. AdaBoost is a classifier that consists of stumps, i.e. one-depth trees. After the construction of each tree, misclassified points are identified and given a heavier weight. Then sequentially, trees are trained on modified data. Let $H_{t-1}$ be our booster classifier at step $t-1$. Then, we define a boosted classifier by a linear combination of weak classifiers, giving
\begin{eqnarray}
H_{t-1}(\mathbf{x}_i) = \alpha_1 h_1(\mathbf{x}_i) + \dots + \alpha_{t-1} h_{t-1}(\mathbf{x}_i)\nonumber.
\end{eqnarray}
At step $t$, we add another weak classifier with a weight $\alpha_t$. The updated classifier is then given by
\begin{eqnarray}
\label{eq:update}
H_t(\mathbf{x}_i) = H_{t-1}(\mathbf{x}_i) + \alpha_t h_t(\mathbf{x}_i).
\end{eqnarray} 
Now we aim to find $h_t$ and $\alpha_t$ such that the loss function is minimised. Therefore, we have an optimisation problem, given by
\begin{eqnarray}
\label{eq:optimAda}
\min_{h} \sum_{i=1}^N L(\tilde{y}_i, H_{t-1}(\mathbf{x}_i) + \alpha_t h_t(\mathbf{x}_i)),
\end{eqnarray}
where $\tilde{y} \in \{ -1, +1\}$. In AdaBoost algorithm, the loss function is an exponential loss. Thus, the loss function in \ref{eq:optimAda} at step $t$ can be written as 
\begin{eqnarray}
\label{eq:lossAda}
L_t(h) &=& \sum_{i=1}^N \exp\{ - \tilde{y}_i [H_{t-1}(\mathbf{x}_i) + \alpha_t h_t(\mathbf{x}_i)]\}.
\end{eqnarray}
Solving this optimisation problem, the optimal values of $h$ and $\alpha$ are given by (see \ref{sec:optada} for details)
\begin{itemize}
\item $h_t^{*} = \arg \min_{h} w_{it}\, I(\tilde{y}_{i} \neq h(\mathbf{x}_i))$
\item $\alpha_t^{*} = \frac{1}{2} \log \frac{1 - e_t}{e_t}$, where $e_t$ is given by
\begin{eqnarray}
e_t = \frac{\sum_{i=1}^N w_{it} I(\tilde{y}_i \neq h_t(\mathbf{x}_i))}{\sum_{i=1}^N w_{it}}\nonumber.
\end{eqnarray}
\end{itemize}
We can implement this algorithm using the following steps:
\begin{enumerate}
\item Initialise weights $w_i = 1/N$ and $H_0 =0$
\item Fit a weak classifier to the training set using weights $w_i$
\item Compute $e_t = \frac{\sum_{i=1}^N w_{it} I(\tilde{y}_i \neq h_t(\mathbf{x}_i))}{\sum_{i=1}^N w_{it}}$
\item Compute $\alpha_t = \frac{1}{2} \log \frac{1 - e_t}{e_t}$
\item 
\begin{itemize}
\item Update weights: $w_{it} := w_{it} e^{-\alpha_t \tilde{y}_i h_t(\mathbf{x}_i)}$
\item Re-normalise $w_{it}$
\end{itemize}
\item Update the boosted classifier $H_t(\mathbf{x}) = H_{t-1}(\mathbf{x}) + \alpha_t h_t(\mathbf{x}_i)$
\item The output is $H(\mathbf{x}) = \textnormal{sign} \left(\sum_{t=1}^T \alpha_t h_t\right)$.
\end{enumerate}
See, Murphy (2012) for more details.

\subsection{Logistic regression}
\label{sec:logistic}
Logistic regression is a special form of Generalised Linear Models (GLM) with the conditional probability given by
\begin{eqnarray}
y_i| \mathbf{x} \sim \textnormal{Bernoulli}(h_{\theta}(\mathbf{x}))\nonumber
\end{eqnarray}
where $h_{\theta}(\mathbf{x})$ is our parametric hypothesis. We have
\begin{eqnarray}
h_{\theta}(\mathbf{x}) = g(\mathbf{\theta}^T \mathbf{x}+\theta_0) = \frac{1}{1+e^{-\mathbf{\theta}^T \mathbf{x}-\theta_0}},
\end{eqnarray}
where $g$ is a link function that relates the linear predictor $\mathbf{\theta}^T \mathbf{x}+\theta_0$ to $h$. In a logistic model, the link function is a sigmoid or logit function and is given by
\begin{eqnarray}
\label{eq:sigmoid}
g(x) = \frac{1}{1+e^{-x}}.
\end{eqnarray}
Logistic regression is a parametric model and to estimate the parameters we need to optimise the negative log-likelihood (NLL) function. We can add a penalty term to the NLL function to overcome the problem of over-fitting. 
Our optimisation problem is then given by 
\begin{eqnarray}
\label{eq:NLL}
\min_{\theta} \, \left(- \sum_{i=1}^N [y_i \log h(\mathbf{x}) + (1-y_i) \log(1-h(\mathbf{x}))] + \lambda R(\theta)\right),
\end{eqnarray}
where the hyperparameter $\lambda$ is the regularisation parameter and $R(\theta)$ is the regularisation term that can take the following forms:
\begin{itemize}
\item $L_1$ penalty: $\|\theta\|_1 = \sum_{i=1}^N |\theta_i|$ (Lasso regression)
\item $L_2$ penalty: $\frac{1}{2}\|\theta\|_2^2 = \frac{1}{2}\theta^T\theta = \frac{1}{2} \sum_{i=1}^N \theta_i^2$ (Ridge regression)
\end{itemize}
When $\lambda$ is large, bias is high and variance is low. In this case, most coefficients are very small and close to $0$. On the other hand, when $\lambda$ is small, bias is low and variance is high therefore we have the problem of over-fitting. To solve the optimisation problem, we use SAGA which is a version of stochastic average gradient descent (SAG) (Schmidt et al., 2017) and can be applied to cases without penalty, with $L_1$, and $L_2$ penalty functions (see also Defazio et al., 2014).

\subsection{Support Vector Classifier}
Support Vector Classifiers (SVC) were introduced by Vapnik (1982) and developed by Cortes and Vapnik (1995). They can be used both in linear and non-linear classification problems. Suppose $\theta^T \mathbf{x}+\theta_0 = 0$ is a hyperplane with $\theta$ being the normal vector to the hyperplane. Also, let $y_i \in \{-1,1\}$ be our labels. Then we define our hypothesis by
\begin{eqnarray}
\label{eq:h}
h_{\theta}(x) = g(\theta^T \mathbf{x} + \theta_0) = \textnormal{sign}(\theta^T \mathbf{x} + \theta_0),
\end{eqnarray}
where $g(z) = 1$ if $z \geq 0$, and $g(z) = -1$ if $z<0$. We can add a constraint that for $y = 1$ if we classify our sample correctly, $\theta^T \mathbf{x} + \theta_0 \geq  1$, and for $y = -1$, $\theta^T \mathbf{x} + \theta_0 \leq -1$. In other words, $y_i(\theta^T \mathbf{x}+\theta_0) \geq 1$. Our goal is to maximise the distance between the closest points to the hyperplane. Let $M$ denote this distance, given by
\begin{eqnarray}
M &=& \min_{\{x:y=1\}} \frac{\theta^T \mathbf{x}}{\|\theta\|} - \max_{\{x:y=-1\}} \frac{\theta^T \mathbf{x}}{\|\theta\|}\nonumber\\
&=& \frac{1-\theta_0}{\|\theta\|} - \frac{-1 - \theta_0}{\|\theta\|} = \frac{2}{\|\theta\|}\nonumber,
\end{eqnarray}
where in the first line $\frac{\theta^T \mathbf{x}}{\|\theta\|} $ represents the normal distance from the closest points to the hyperplane. We need to find $\theta$ and $\theta_0$ such that $M$ is maximised. Hence
\begin{eqnarray}
\label{eq:opt3}
\min_{\theta, \theta_0} && \frac{1}{2} \|\theta\|^2 \nonumber\\
s.t.&& y_i (\theta^T \mathbf{x}_i + \theta_0) \geq 1, \quad i = 1, \dots, N.
\end{eqnarray}
This is called a hard margin and may lead to over-fitting. We can also introduce a soft margin, where some instances like $\xi_i$ are either misclassified or are located within the margin boundary. In this case, the objective function includes $L_1$ regularisation and \ref{eq:opt3} is modified as 
\begin{eqnarray}
\label{eq:svcreg}
\min_{\theta, \theta_0} && \frac{1}{2} \|\theta\|^2 + C \sum_{i=1}^N \xi_i\nonumber\\
s.t. && y_i (\theta^T \mathbf{x}_i + \theta_0) \geq 1 - \xi_i, \quad i = 1, \dots, N,\nonumber\\
&& \xi_i \geq 0, \quad i = 1, \dots, N.
\end{eqnarray}
To solve this problem, we can apply the Lagrangian multiplier technique (see \ref{sec:lag} for details). The optimal value of $\theta^{*}$ is then given by
\begin{eqnarray}
\label{eq:sol}
\theta^{*} = \sum_{i=1}^N \lambda_i y_i \mathbf{x}_i,
\end{eqnarray}
where $\lambda$ is the Lagrangian multiplier. To predict the class of an instance, we need to substitute $\theta^{*}$ in \ref{eq:h}, giving
\begin{eqnarray}
\theta^{*T} \mathbf{x} + \theta_0 = \left(\sum_{i=1}^N \lambda_i y_i \mathbf{x}_i \right)^T \mathbf{x} + \theta_0 =  \sum_{i=1}^N \lambda_i y_i \langle \mathbf{x}_i, \mathbf{x}\rangle + \theta_0. 
\end{eqnarray}
Further, using Lagrangian multiplier technique, we can write \ref{eq:svcreg} as (see \ref{sec:lag} for details)
\begin{eqnarray}
\label{eq:svcLag}
\max_{\lambda} && \sum_{i=1}^N \lambda_i - \frac{1}{2} \sum_{i=1}^N \sum_{i=1}^N \lambda_i \lambda_j y_i y_j \langle \mathbf{x}_i, \mathbf{x}_j \rangle\nonumber\\
s.t. && 0 \leq \lambda_i \leq C, \quad i = 1, \dots, N\nonumber\\
&& \sum_{i=1}^N \lambda_i y_i = 0.
\end{eqnarray}  
 As we can see both the objective function in \ref{eq:svcLag} and the optimal value of $\theta^{*}$ in \ref{eq:sol} only depend on the inner product of the features.  At the beginning of this section, we explained that SVC can be applied to non-linear classification. This is possible through kernels. Let $\kappa(x, x')$ be a kernel function, which is a real-valued, positive and symmetric function. Then, \ref{eq:svcLag} can be kernelised by setting $\kappa(x, x') = \langle x, x' \rangle$. Here, we use radial basis function (rbf) $\exp\left(- \gamma \|x - x'\|^2\right)$ and apply LIBSVM to implement SVC. This is a library developed by Chang and Lin in 2000  and can handle multi-class classification problems (Chang and Lin, 2022).

\subsection{Neural Networks (NN)}
\label{sec:NN}
NNs also known as multilayer Perceptron (MLP) consist of an input layer, one or more hidden layers, and one output layer. A NN with a large number of hidden layers is called a deep NN (DNN). In a NN, each layer has a series of units called neurons that model the neurons in a biological brain. Data, $\mathbf{x}$, move from the input units which are usually equal to the number of features through the hidden units and come out from the output units. Different layers perform different transformations on their inputs through an activation function $g$, which can take different forms such as sigmoid, $\tanh(x) = (e^{x}-e^{-x})/(e^{x}+e^{-x})$, rectified linear unit, ReLU$(x) = \max(0,x)$, etc. When all the units in a layer are connected to every unit in the previous layer, the layer is called a fully connected or a dense layer. Each input to a unit has an associated weight $\theta$ which can be adjusted through the learning process, known as backpropagation algorithm, introduced by Rumelhart et al. (1986). This method applies gradient descent to estimate the parameters $\theta$ using automatic differentiation to compute the gradients automatically through two passes - forward and backward. The forward pass for a three-layer (two hidden layers and one output layer) fully connected NN with $j = 1, \dots, m^{[L]}$ hidden units in layer $L$ is given by
\begin{itemize}
\item $z^{[1]}_j = \theta^{[1]^{T}}_j \mathbf{x} + \theta_{0j}^{[1]};\quad  a^{[1]}_j = g(z^{[1]}_j)$, where $\theta^{[1]}_j \in R^d$ and $\mathbf{a}^{[1]} = [a_1^{[1]}, \dots, a^{[1]}_{m^{[1]}} ]^T \in R^{m^{[1]}}$
\item $z_j^{[2]} = \theta^{[2]^{T}}_j \mathbf{a}^{[1]} + \theta_{0j}^{[2]};\quad  a_j^{[2]} = g(z^{[2]})$, where $\theta_j^{[2]} \in R^{m^{[1]}}$ and  $\mathbf{a}^{[2]} = [a_1^{[2]}, \dots, a^{[2]}_{m^{[2]}} ]^T \in R^{m^{[2]}}$
\item $z^{[3]} = \theta^{[3]^{T}} \mathbf{a}^{[2]} + \theta_{0}^{[3]};\quad  a^{[3]} = g(z^{[3]})$, where $\theta^{[3]} \in R^{m^{[2]}}$,
\end{itemize} 
where $\theta^{[L]}_j$ is the associated weight to unit $j$, layer $L$ and $\theta^{[L]}_{0j}$ is the bias in layer $L$. As we can see when $g$ is a sigmoid function, a NN with one unit performs similar to a logistic model. In forwardpropagation, the inputs to layer 1 are data, $\mathbf{x}$, and the output is $z^{[1]}$ which is transformed through the activation function $g$. Then, the input to layer 2, is $\mathbf{a}^{[1]}$ which is a vector of dimension $m^{[1]}$, and the output is $\mathbf{a}^{[2]}$ which is a vector of dimension $m^{[2]}$. The output layer, which in the case of binary classification has only one unit, transforms $z^{[3]}$ and outputs $h_{\theta}(\mathbf{x}) = a^{[3]}$. At this stage, the output is compared with the actual label through a loss function $L(h_{\theta}(\mathbf{x}), y)$. Then, the total error, $E$, of the network over all layers and units is calculated. To minimise the network's error by gradient descent, the partial derivative of the loss function with respect to each parameter is needed. Backward pass applies the chain rule to compute these partial derivatives and applies the gradient descent to update the parameters (see, e.g., Bishop, 2005, and G\'{e}ron, 2019). For layer $L$ the updated parameters are given by
\begin{eqnarray}
\theta^{[L]} := \theta^{[L]} - \alpha \frac{\partial E}{\partial \theta^{[L]}} \nonumber\\
\theta_0^{[L]} := \theta_0^{[L]} - \alpha \frac{\partial E}{\partial \theta_0^{[L]}}
\end{eqnarray} 
where $\alpha$ is the learning rate, which is a hyperparameter. One problem with the gradient descent is that it can be slow to reach the minimum and if we increase the learning rate, it may overshoot. There are different variants of gradient descent. One popular method is the adaptive method with momentum (adam), introduced by Kingma and Ba (2015). It is given by
\begin{eqnarray}
\theta := \theta - \alpha \frac{\mathbf{\hat{v}}}{\epsilon + \sqrt{\mathbf{\hat{s}}}},
\end{eqnarray}
where
\begin{itemize}
\item $\mathbf{v} := \gamma_{v} \mathbf{v} + (1-\gamma_{v}) d\mathbf{\theta};\quad \mathbf{\hat{v}} = \mathbf{v} / (1-\gamma_v)$
\item $\mathbf{s} := \gamma_s \mathbf{s} + (1-\gamma_s) d\mathbf{\theta}^2; \quad \mathbf{\hat{s}} = \mathbf{s} / (1- \gamma_s)$.
\end{itemize}
When we initialise $v$ and $s$ to zero at the initial steps $v$ and $s$ are close to $0$. To correct this bias, we divide $v$ and $s$ by the correction terms $1-\gamma_v$ and $1- \gamma_s$, respectively. The hyperparameters $\alpha$, $\gamma_{v}$, $\gamma_s$, and $\epsilon$ are usually set to $0.001$, $0.9$, $0.999$, and $1 \times 10^{-8}$, respectively. (See \ref{sec:gradient} for details). 

\par The gradient descent can be implemented in mini-batches. When the size of training data is very large, it is costly to update all parameters by passing the whole sample. We can divide our sample into mini-batches and update our parameters by passing each mini-batch, hence the name mini-batch gradient descent. 

\subsection{Classification metrics}
\label{sec:cross}
A confusion matrix is a tool in classification analysis. Table \ref{tab:modes} shows a confusion matrix and different measures that can be obtained from that. We can obtain the following information from a confusion matrix:
\begin{itemize}
\item Accuracy: $\frac{TP+TN}{TP+TN+FP+FN}$. This measures the overall performance of an algorithm. 
\item Balanced accuracy: $\frac{1}{2} \left(\frac{TP}{TP+FN}+\frac{TN}{TN+FP}\right)$. This measures the average accuracy from both positive and negative classes.
\item Misclassification rate: $\frac{FP+FN}{TP+TN+FP+FN}$. (1-Accuracy).
\item Sensitivity (recall): $\frac{TP}{TP+FN}$. This measures the performance of an algorithm in predicting positive classes. 
\item Precision: $\frac{TP}{TP+FP}$. This measures quality of prediction. 
\item Specificity: $\frac{TN}{TN+FP}$. This measures the performance of an algorithm in predicting negative classes. 
\item Receiver operating characteristics (ROC): It plots sensitivity against specificity across different thresholds. 
\item AUC: it measures the area under ROC curve. 
\end{itemize}
For more details on the topics in this section see Hastie et al. (2009) and Murphy (2012).

\begin{table}{}
\caption{Confusion matrix: positive class (non-responses, $y=1$), negative class (responses, $y=0$)}
\centering
\footnotesize
\begin{tabular}{ccc}\toprule
				& Predicted positive 	(1)						& Predicted negative	 (0) \\\midrule
Actual positive     	& TP	(correctly predicted as positive)				& FN	 (incorrectly predicted as negative)	 \\
Actual negative		& FP (incorrectly predicted as positive)			& TN	 (correctly predicted as negative)	 \\\bottomrule
\end{tabular}
\label{tab:modes}
\end{table}

\section{Results and discussion}
\label{sec:results}
In this section, we explain how we can apply the models in Section \ref{sec:models} to our dataset. After setting column \textit{id} as an index, we have a dataset with $5,820$ rows and $50$ columns (features). All our features except \textit{age} are categorical. Most ML models only accept numerical inputs. First, we separate our features from our class. Then, we pre-process our data. We can encode categorical features using \textit{OrdinalEncoder}. Some objective functions assume that features are centred around zero. We can use \textit{StandardScaler}, $(X-\mu)/\sigma$, where $\mu$ and $\sigma$ are mean and standard deviation, respectively, or \textit{MinMaxScaler}, $(X-\min(X))/(\max(X)-\min(X))$, (suitable for KNN algorithm) to scale numerical features. Further, we use cross-validation for hyperparameter tuning. We split our data into a training set to train our models and a validation set to evaluate our models. We can repeat this, say, 5 times. Therefore, the scaling parameters may be different when a new training and validation set is created. To sequentially combine \textit{transformers}, such as encoders and scalers, and \textit{estimators} such as logistic regression, RF, etc, we can use \textit{Pipeline}. This way, we will not have the problem of data leak, where information outside a training set is used to train a model. After tuning hyperparameters, we split our dataset into training and test set and apply our model to the test set. Then, we evaluate our models using the classification metrics and use the feature importance technique for interpretation. In this method, we permute one feature and look at its impact on the accuracy of our model. The idea is that if a feature is significant, a change in that feature will lead to a change in the performance of the model.  

\par
\begin{figure}[]
\centering
\includegraphics[scale = 0.3]{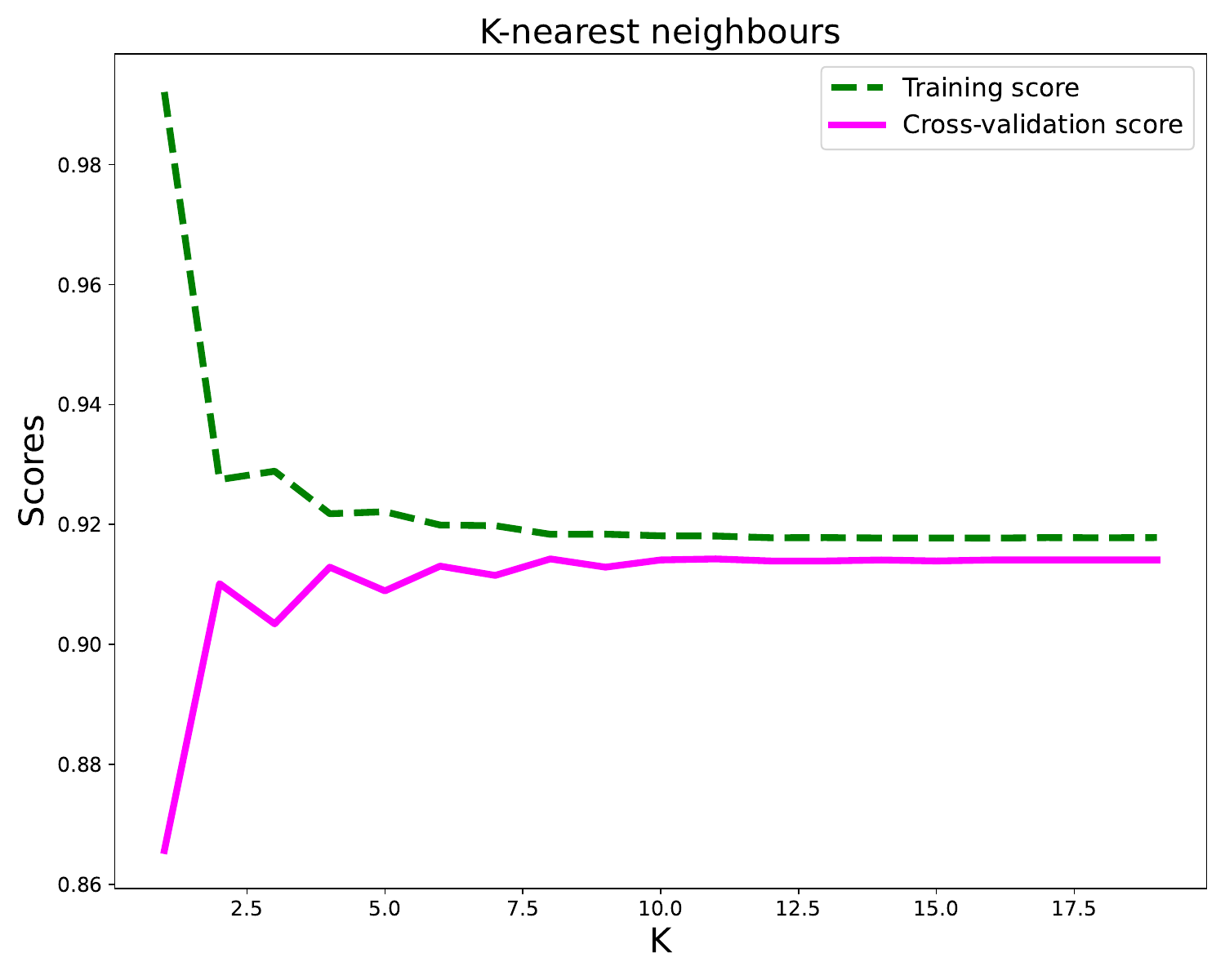}
\includegraphics[scale = 0.3]{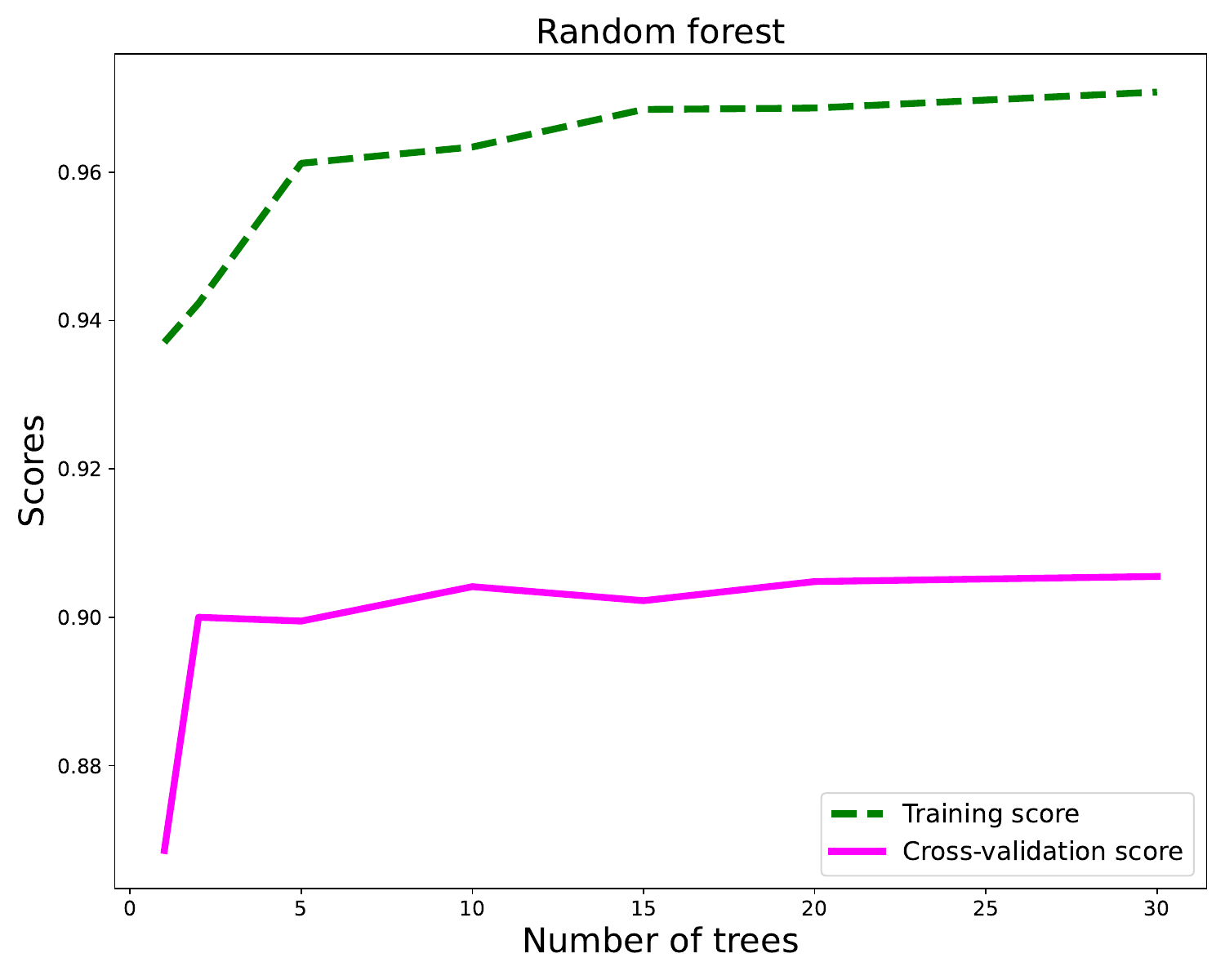}
\includegraphics[scale = 0.3]{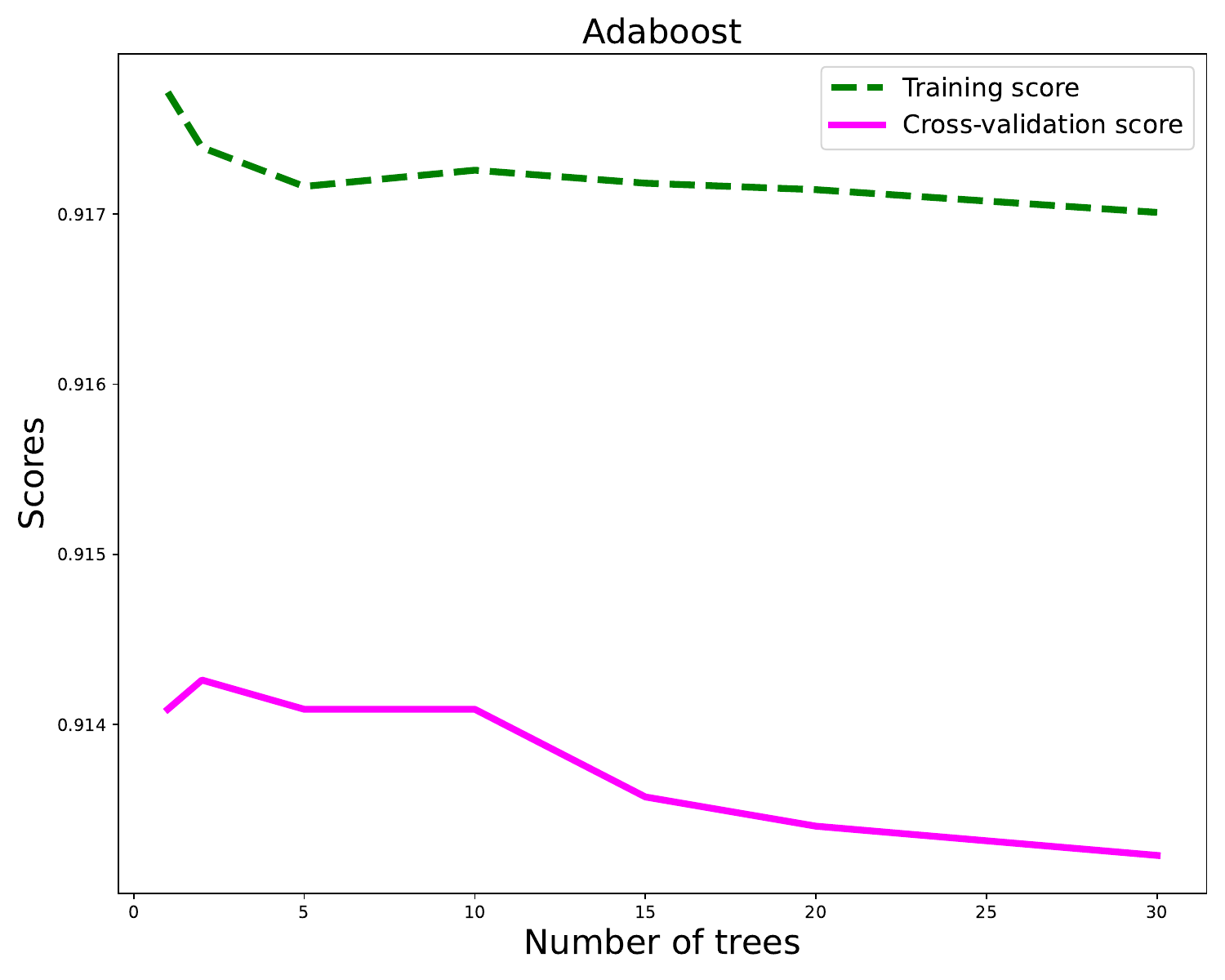}
\includegraphics[scale = 0.3]{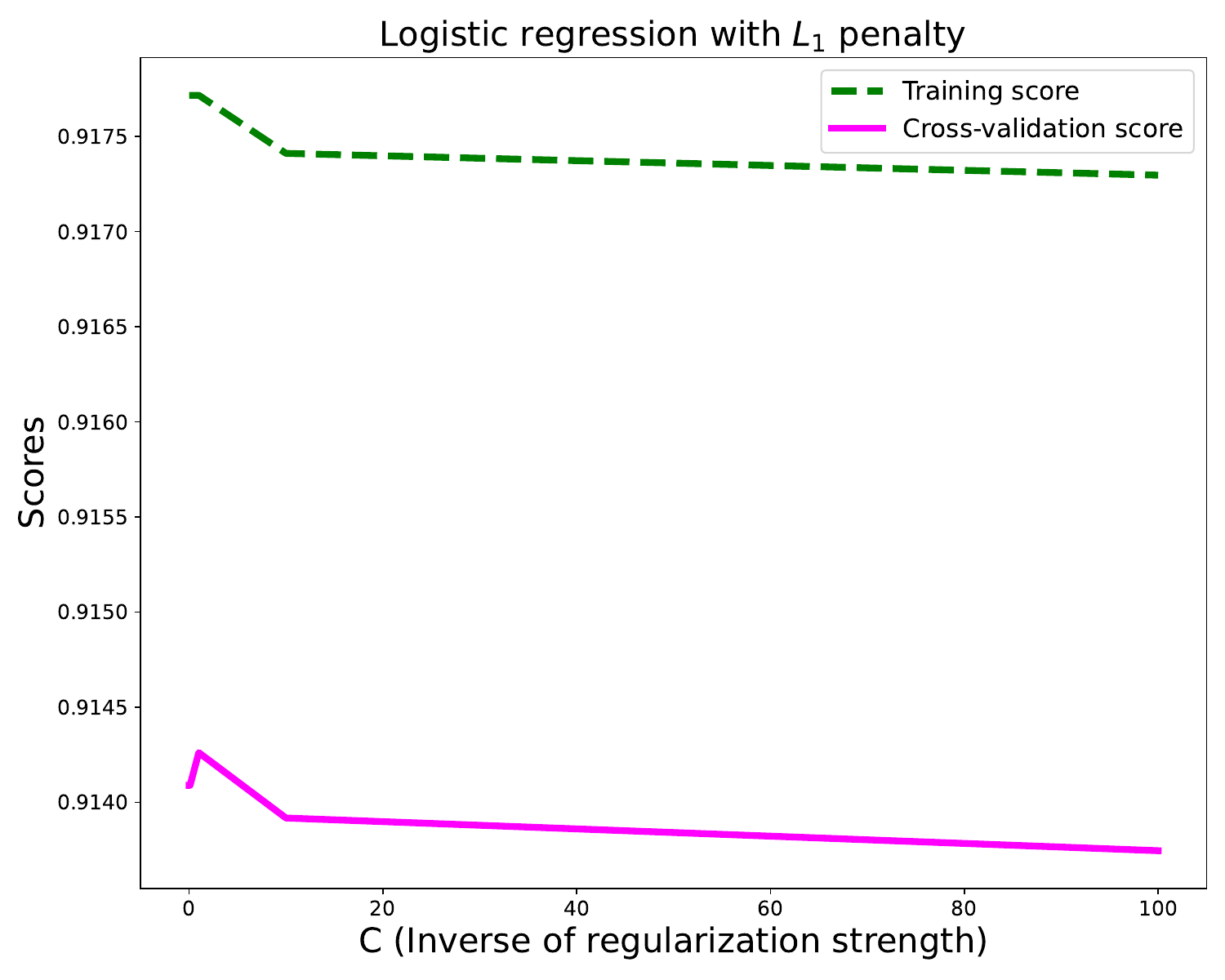}
\includegraphics[scale = 0.3]{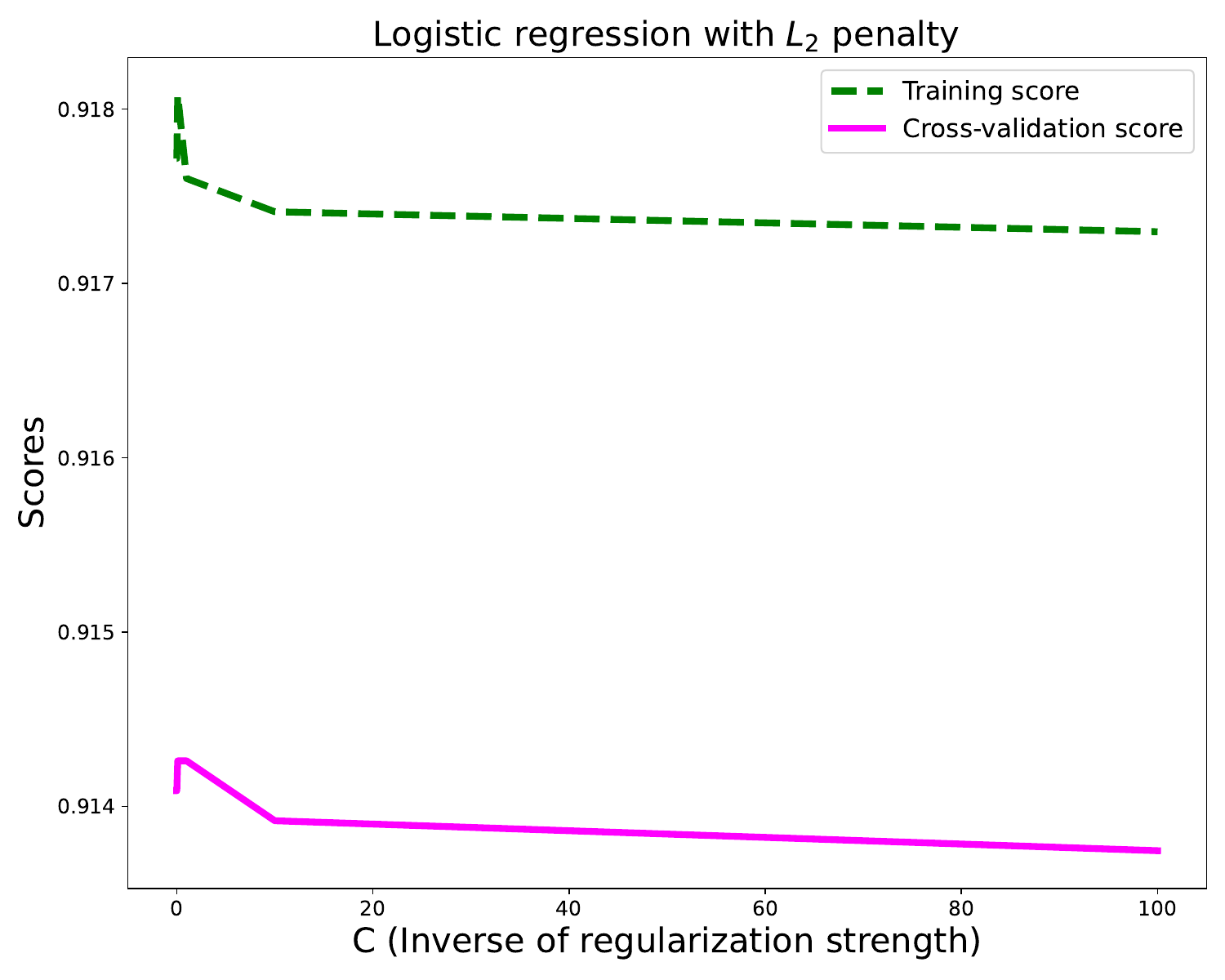}
\caption{Validation curves}
\label{fig:Valid}
\end{figure}

 Figure \ref{fig:Valid} shows the validation curves. To tune hyperparameters, we use \textit{validation\_curve} and apply \textit{ShuffleSplit} as the cross-validation splitting strategy to shuffle our dataset at each iteration. We can see that on the top-left as $K$ increases, training accuracy decreases, and validation accuracy increases. A high value of $K$ makes the algorithm less complex and it will be easier to generalise to an unseen dataset. We select $K=10$ as we have the highest level of validation accuracy at this level. To apply RF and AdaBoost we use \textit{gini} to measure the quality of a split and create fully-grown trees. To build an AdaBoost classifier, we set a decision tree with maximum depth $= 1$ as the base estimator. Scikit-learn uses stagewise additive modelling with a multi-class exponential loss function (SAMME) and (SAMME.R) to implement AdaBoost. These two algorithms were introduced by Hastie et al. (2009a) and unlike the algorithm that we discussed in Section \ref{sec:ensemble}, they can handle multi-class classification. The difference between SAMME and SAMME.R is that the former computes empirical error at step 3 (see Section \ref{sec:ensemble}), whereas the latter uses weighted probability estimates derived from the classifier and converges faster. In Figure \ref{fig:Valid} we can see that the highest level of validation accuracy is achieved when the number of trees is $10$ for RF and $3$ for AdaBoost. The hyperparameter for logistic regression with a penalty function is $C$. To implement logistic regression, we select \textit{saga} and set the maximum number of iterations to $4,000$ for logistic regression with a penalty and $3,000$ without a penalty. We can see in Figure \ref{fig:Valid} that the highest validation accuracy is obtained when $C=1$ for both $L_1$ and $L_2$ penalties. For SVC, we use \textit{GridSearchCV} as we need to select different parameters such as $C$, \textit{kernel}, and the parameters of kernels. Therefore, we need a grid to find a combination of the parameters that gives the highest accuracy. The output in terms of the highest accuracy is an SVC with an rbf kernel, $C=1$ and $\gamma = 0.1$. The last model that we train is a NN. We use \textit{MPLClassifer} in scikit-learn. Although scikit-learn is not a famous library for deep learning, it works well for a vanilla NN and our tabular data. Besides that this allows us to be consistent with other models in terms of pre-processing and evaluation. We apply \textit{GridSearchCV} to select a suitable number of hidden layers and units. We train a 4-layer NN with $32, 16$, and $8$ units, a 3-layer NN with $4$, and $2$ units, and three 2-layer NNs with $50$ units, $4$ units, and $2$ units. We found that a 3-layer NN with $4, 2$ units, tanh activation function (for hidden layers), 1000 epochs (iterations over the whole training data), and adam optimisation method is the best model. We did not change the parameters of adam or the size of batches.

\subsection{Model evaluation and interpretation}
\begin{figure}[]
\centering
\includegraphics[scale = 0.5]{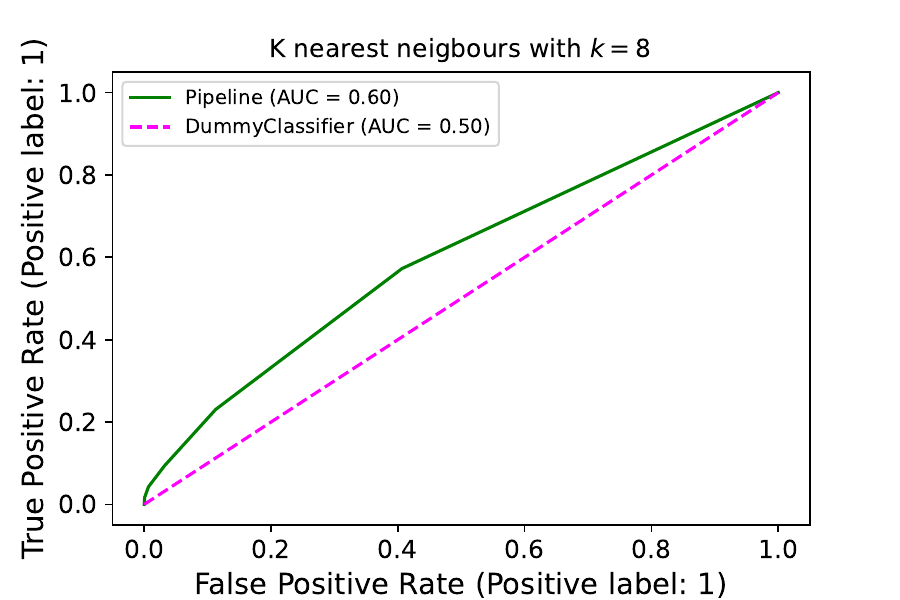}
\includegraphics[scale = 0.5]{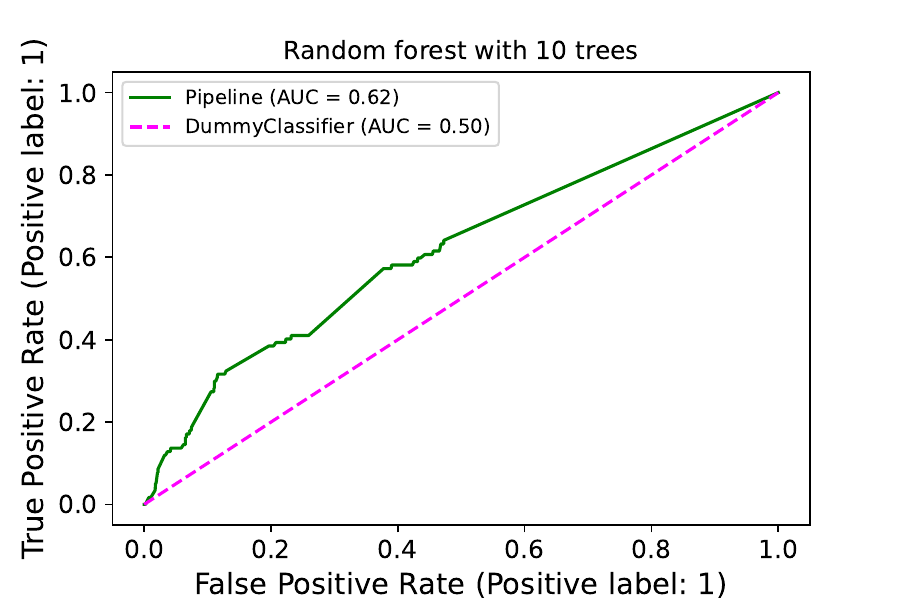}
\includegraphics[scale = 0.5]{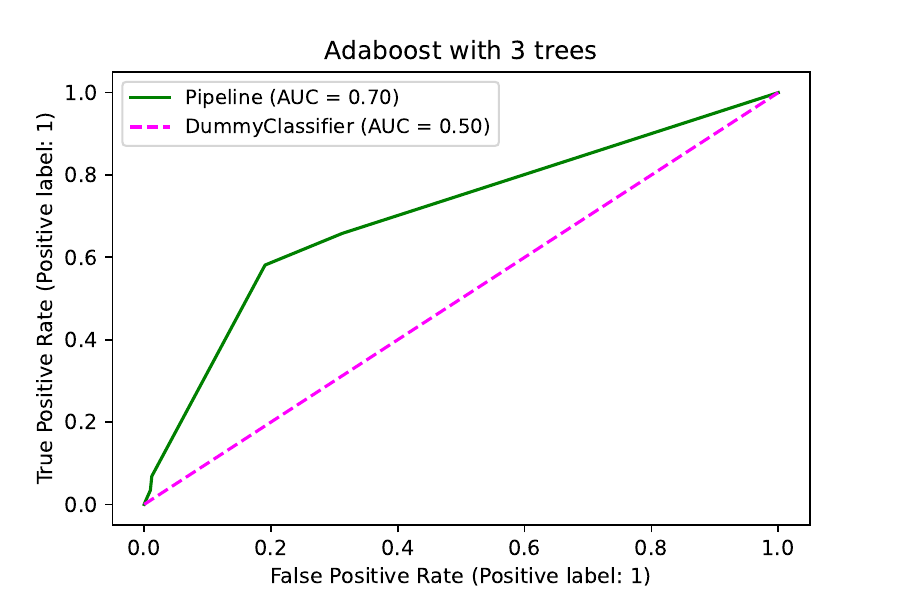}
\includegraphics[scale = 0.5]{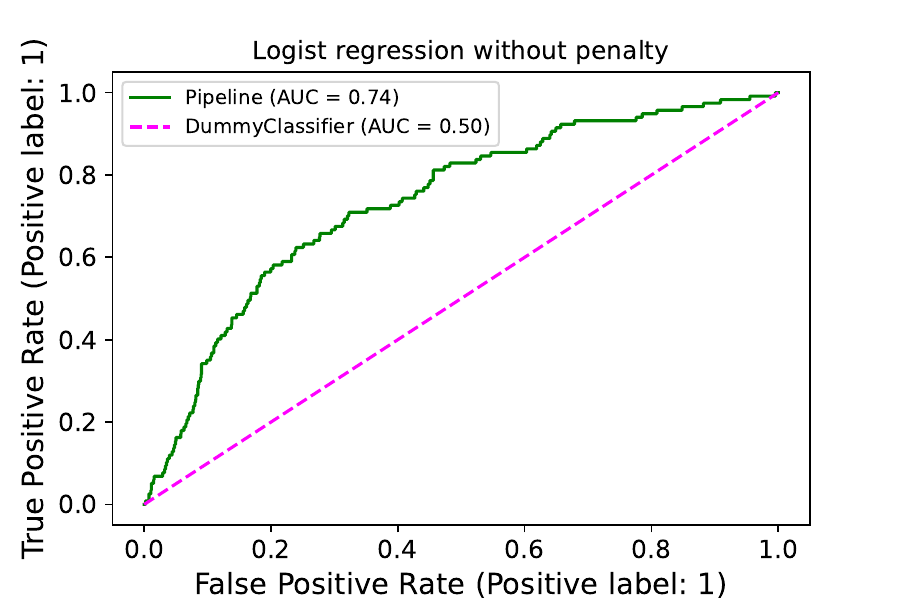}
\includegraphics[scale = 0.5]{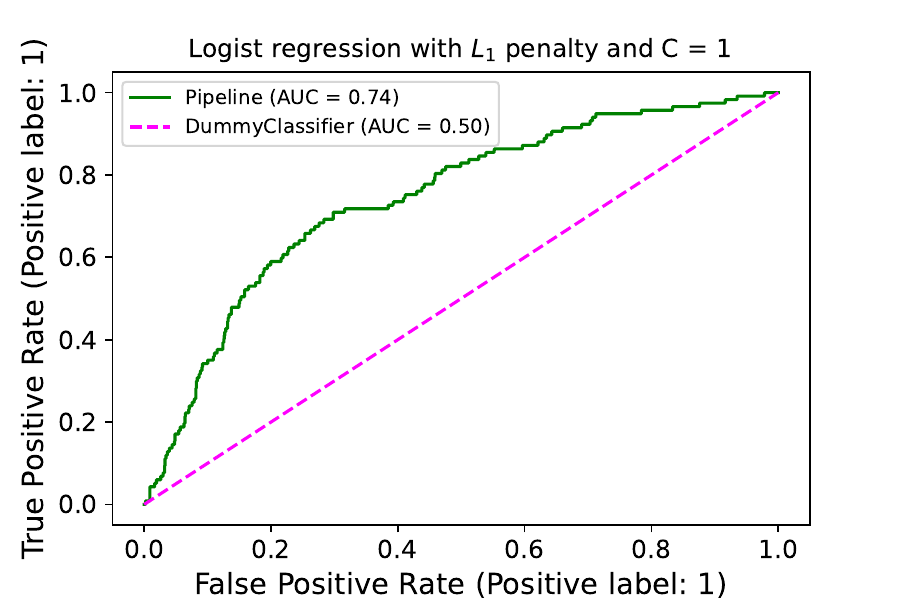}
\includegraphics[scale = 0.5]{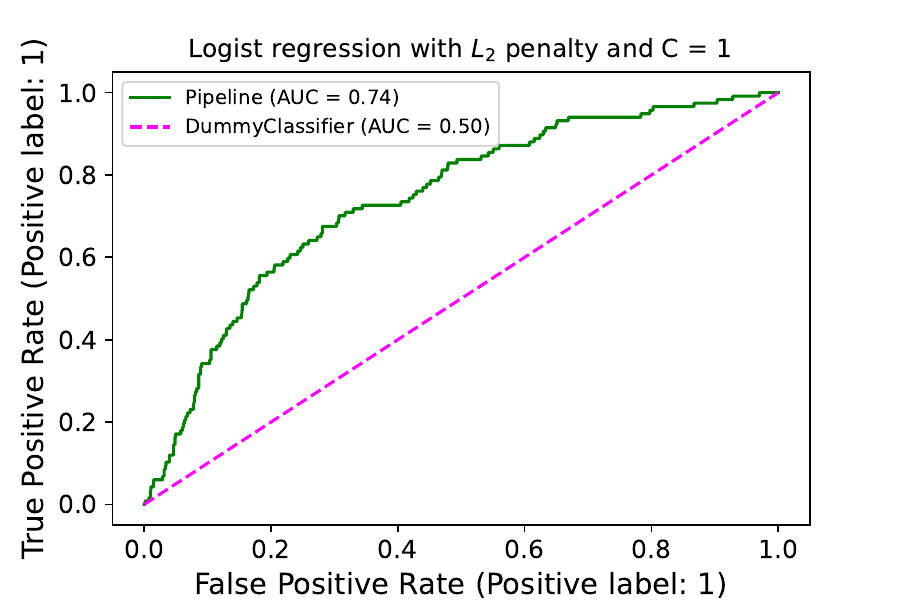}
\includegraphics[scale = 0.5]{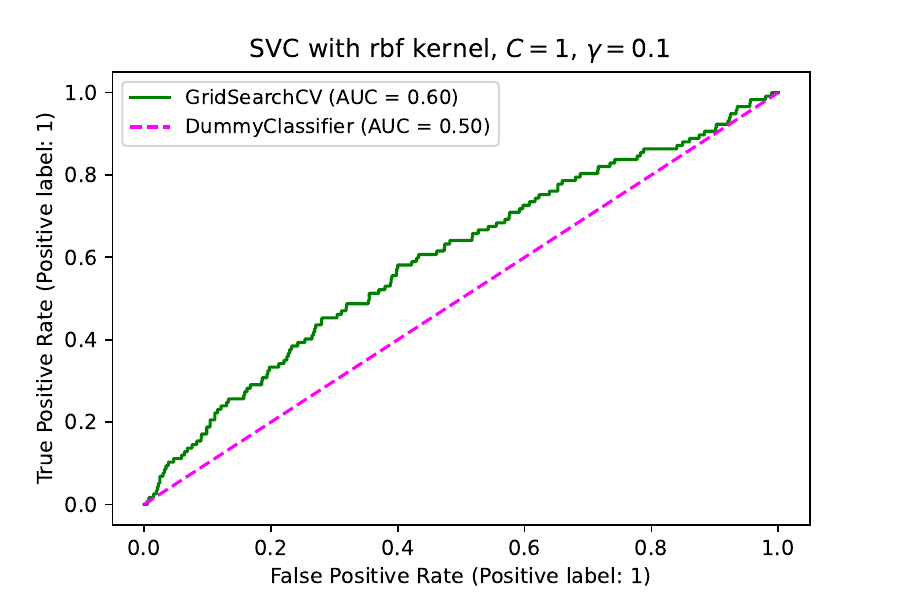}
\includegraphics[scale = 0.5]{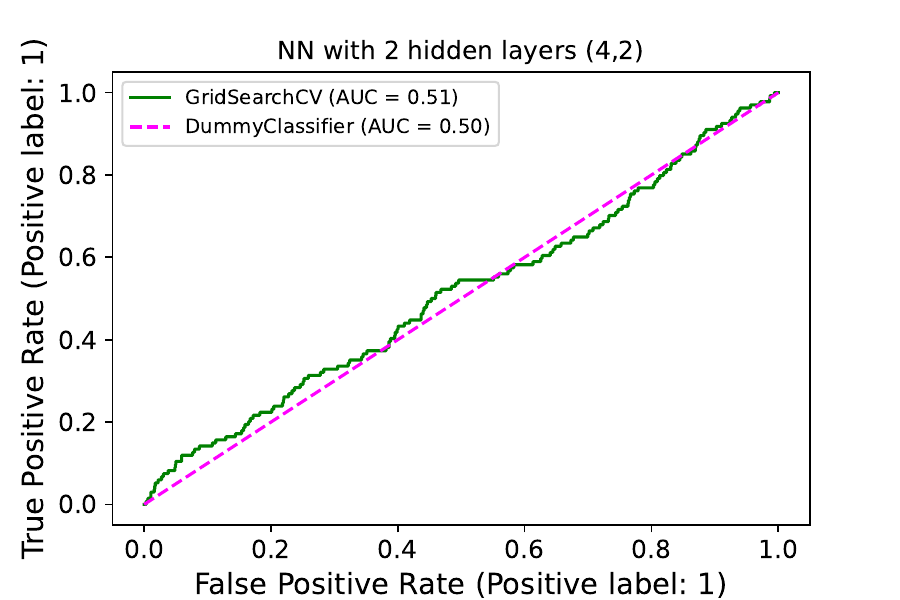}
\caption{ROC curves}
\label{fig:ROC}
\end{figure}

Tables \ref{tab:Log} and \ref{tab:metrics} present the classification metrics for our models. As we can see, RF outperforms other models in terms of training accuracy with NN coming after that. However, it does not perform well in the case of an unseen dataset. KNN and SVC have the best performance in terms of test accuracy. When the number of the positive class is much less than the number of negative class, we have a problem of unbalanced dataset. In this case, accuracy is not a good measure of the model's performance. Instead, we can use balanced accuracy. All models perform relatively similar in terms of balanced accuracy and the highest score is obtained by RF. TP and TN show the number of cases that are correctly classified by a model. Our goal is to predict the number of the minority class. As we can see the highest number of positive class, which is predicted correctly, is given by RF and SVC and NN do not predict any cases from the positive class. FP and FN show misclassification. However, sometimes it is more costly to misclassify the positive class. Here, all models except NN perform similarly in terms of FN. RF misclassifies the least number of positive cases as negative. The highest level of precision and recall are obtained by KNN and RF, respectively. In terms of generalisation which is represented by AUC, logistic regressions outperform other models. These models are used by UK ONS (ONS, 2009) and US Bureau (Erdman and Bates, 2017) to predict the non-response rate. In Figure \ref{fig:ROC}, we can see that logistics regressions with penalty perform even better than a logistic regression without penalty, which is in line with our expectation as models with penalty functions are less complex and more generalisable. 
       
\par One problem with ML models is interpretability. Here, we use \textit{permutation\_importance} to identify the significant features that affect responses. 
Figures in Appendix \ref{app:ada} show the significant features selected by models and algorithms. Some of those features, in the same order, are also presented in Tables \ref{tab:Log} and \ref{tab:metrics}. The mode of interview is found to be a significant feature, although in a different order, by all models except SVC. In Figure \ref{fig:interview} we can also see that the proportion of participants who did not participate in wave 2 and were interviewed online is about $5\%$, whereas for the other mode of the interview is $20\%$. Modes of survey as one of the factors that affect the response rate have been considered by Cho et al. (2021) and Dutwin and Buskirk (2022). Cho et al. (2021) show that traditional methods of survey lead to decreasing response rate and hence new modes of survey should be introduced. If we assume that COVID-related health issues have affected the response rate, we can see that some of these factors have been selected by our models. During the pandemic, some regions were more affected by COVID-19. We can see that region and rural/urban are two significant factors adopted by most models except AdaBoost. This is in line with the available statistics provided by ONS \footnote{\url{https://www.ons.gov.uk/peoplepopulationandcommunity/healthandsocialcare/healthandlifeexpectancies/bulletins/healthstatelifeexpectanciesuk/2018to2020} Accessed Decemeber 2022} over the period 2018-2020. See, also, McGowan and Bambra (2022) and the references therein. Physical activity is another significant factor chosen by KNN, RF, and SVC. In Figure \ref{fig:activity} we can see that a large proportion of people who did not participate in wave 2 had about the same or less than usual physical activities. There is evidence that physical activities have affected mental health during COVID-19. See, e.g., Pears et al. (2022), and Marconcin et al. (2022) among others. We can also see some other COVID-19-related factors selected by some models. For example, the COVID-19 test is selected by KNN and SVC, and shortness of breath by logistic regressions with a penalty. Another factor selected by RF, SVC, NN, and logistic regression without a penalty, is the employment status (pre-COVID) and during the pandemic. This result agrees with D'Angel et al. (2022) that the loss of jobs had a more severe impact on the health of middle-aged people during the pandemic.      

\begin{table}{}
\caption{Classification metrics for Logistic regression with and without penalty and NN}
\centering
\scriptsize
\begin{tabular}{lcccc}\toprule
 				& $L_1$ penalty with $C=1$			& $L_2$ penalty with $C=1$				& Without penalty 	& 3-layer NN with $(4,2)$ units\\\midrule
Training accuracy 	& 0.9173							& 0.9173								& 0.9168			& 0.9207	\\
Test accuracy		& 0.9175							& 0.9175								& 0.9168			& 0.9079	\\
Balanced accuracy	& 0.5030							& 0.5030								& 0.5020			& 0.5000	\\
Classification error	& 0.0825							& 0.0825								& 0.0832			& 0.0921	\\
TP				& 1								& 1									& 1				& 0		\\
TN				& 1,334							& 1,334								& 1,333			& 1,321	\\
FP				& 4								& 4									& 5				& 0		\\
FN				& 116							& 116								& 116			& 134	\\
Precision			& 0.2000							& 0.2000								& 0.1667			& 		\\
Recall(Tpr)		& 0.0085							& 0.0085								& 0.0085			& 0		\\
Fpr				& 0.0030							& 0.0030								& 0.0037			& 0		\\
Specificity			& 0.9970							& 0.9970								& 0.9963			& 1		\\
AUC				& 0.74							& 0.74								&0.74			& 0.51	\\\midrule
Significant features	& Cohort							& Cohort								& Cohort			& Region	\\
				& Age							& Age								& Age			& High blood pressure\\
				& Interview mode					& Interview mode						& Interview mode	& Employment\\
				& Dementia						& Dementia							& Dementia		& Employment (pre-COVID)\\
				& Disability						& Breath short							& Ear disease		& Interview mode \\
				& Breath short						& Arthritis								& Employment 		& \\
				& Rural/urban						& Disability							& Rural/urban		& \\\bottomrule	
\label{tab:Log}								
\end{tabular}
\end{table}

\begin{table}{}
\caption{Classification metrics for KNN, RF, AdaBoost and SCV}
\centering
\scriptsize
\begin{tabular}{lcccc}\toprule
 				& KNN with $K = 8$ 				&  RF with 10 trees				& AdaBoost with 3 trees		& SVC with rbf, $C=1$ and $\gamma=0.1$ \\\midrule
Training accuracy 	& 0.9173							& 0.9638						& 0.9166			& 0.9168\\
Test accuracy		& 0.9203							& 0.9065						& 0.9182			& 0.9196	\\
Balanced accuracy	& 0.5080							& 0.5320						& 0.5030			& 0.5000\\
Classification error	& 0.0797							& 0.0935						& 0.0818			& 0.0804	\\
TP				& 2								& 10							&1				& 0	\\
TN				& 1,337							& 1,309						& 1,335			& 1,338		\\
FP				& 1								& 29							& 3				& 0		\\
FN				& 115							& 107						&116				& 117	\\
Precision			& 0.667							& 0.2564						& 0.2500			& 	\\
Recall(Tpr)		& 0.0171							& 0.0855						& 0.0085			& 0		\\
Fpr				& 0.0007							& 0.0217						& 0.0022			& 0	\\
Specificity			& 0.9930							& 0.9783						& 0.9978			& 1	\\
AUC				& 0.60						 	& 0.62						& 0.70			& 0.60		\\\midrule
Significant features	& Region							& Region						& Dementia		& Region\\
				& Age							& Interview mode				& Interview mode	& Living place\\
				& Interview mode					& Cohort 						&				& COVID test\\
				& CV test							& Physical activity				&				& Employment (pre-COVID)\\
				& Physical activity					& Disability					&				& Employment  	\\
				& Rural/urban						& Rural/urban					&				& Physical activity\\ 
				& Fatigue							&Employment 					&				& Rural/urban\\ \bottomrule	
\label{tab:metrics}								
\end{tabular}
\end{table}

\subsection{Discussion}
\begin{figure}[H]
\centering
\includegraphics[scale = 0.30]{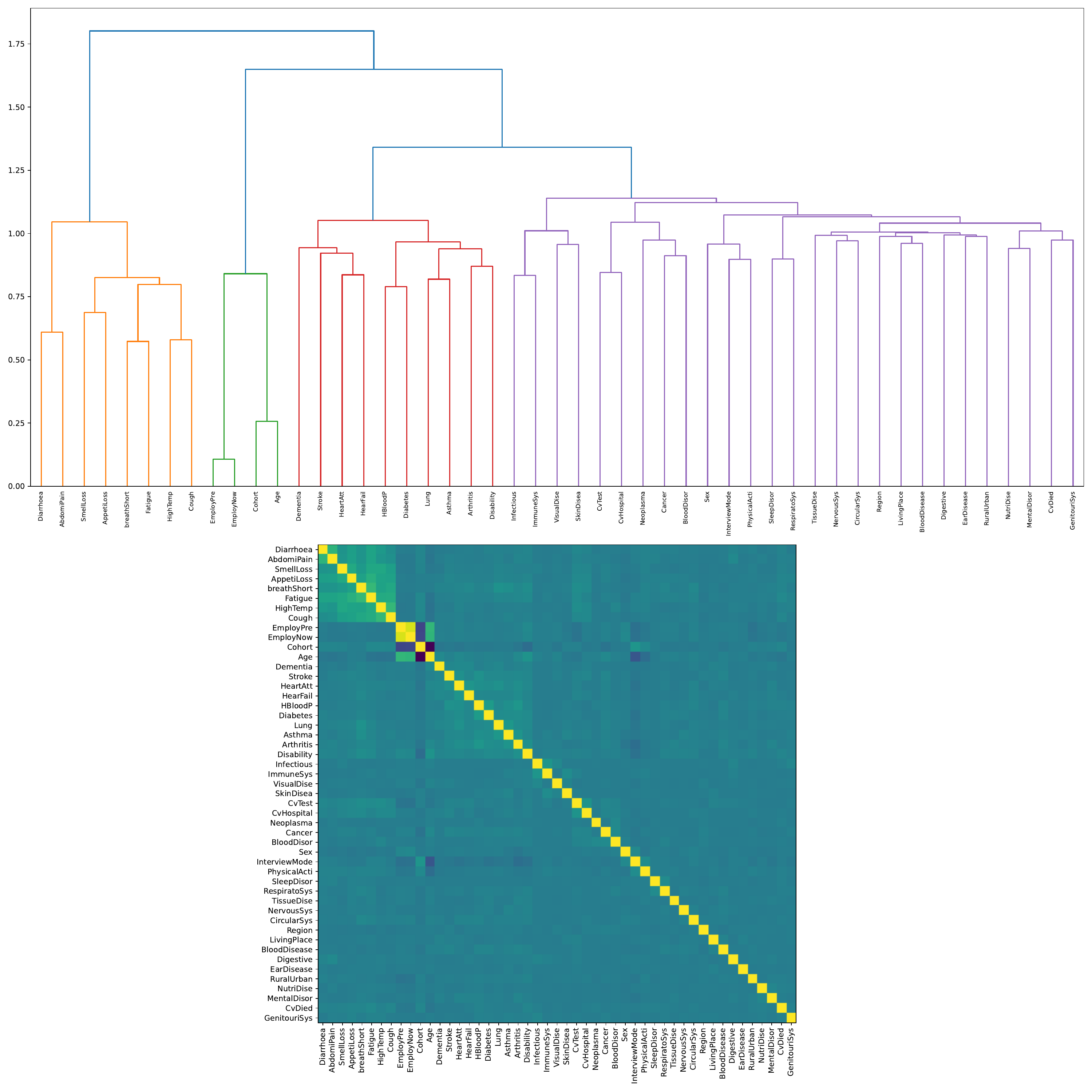}
\caption{Heatmap and the relationship between features}
\label{fig:dendo}
\end{figure}

In the previous section, we saw that the best accuracy that we got was about $92\%$ which is not a very good outcome given that the accuracy of a null model that only considers the most frequent class is $91.06\%$. Our dataset is highly imbalanced and an accuracy score for an imbalanced dataset is misleading when our goal is to predict the minority class. We also reported other measures like balanced accuracy which is the average of sensitivity and specificity. We did some other experiments with our dataset which we did not report here. In the previous section, we used \textit{OrdinalEncoder}. This encoder assigns discrete values in alphabetical order. We also tried \textit{OneHotEncoder} which considers each category of a feature as a binary variable. For example, with a feature with $6$ categories, we will have $5$ new features (dropping one category to avoid collinearity). This did not help with improving the performance of our models. Looking at the graphs of feature importance, we can see a small change in accuracy after the permutation of features. This suggests that those features are not important. However, if we look at Figure \ref{fig:dendo} we can see a heatmap that shows the correlation between our features and a dendrogram that shows the clusters of our features. We then selected one feature from each cluster arbitrarily and trained our models based on those features. We observed small changes in accuracy. Ramyachitra and Manikandan (2014) provide a review of imbalanced datasets and the techniques such as over-sampling and under-sampling of classes to deal with it. Imbalanced datasets are common in areas such as fraud detection, medical diagnostics, churn prediction, etc, and dealing with this type of dataset is challenging.

\section{Conclusion}
In this study, we applied ML models to predict the non-responses in the ELSA COVID-19 dataset and identified the factors that affect non-responses. We analysed this dataset and found that logistic regressions, which are also used by UK ONS and US Bureau, are good at generalisation. However, KNN and RF outperform other models in terms of test and imbalanced accuracy, respectively. We also showed that significant factors selected by our models are in agreement with statistics and literature.  
The prediction of non-responses in a survey is a binary classification problem. Therefore, our methodology can be applied to any other classification problems such as policy renewal in the insurance context.  

\newpage

\appendix

\section{Appendix}
\label{app:ada}

\setcounter{table}{0}
\setcounter{figure}{0}
\setcounter{equation}{0}
\renewcommand{\thetable}{A\arabic{table}}
\renewcommand{\thefigure}{A\arabic{figure}}
\renewcommand{\theequation}{A\arabic{equation}}

\begin{figure}[H]
\centering
\includegraphics[scale = 0.3]{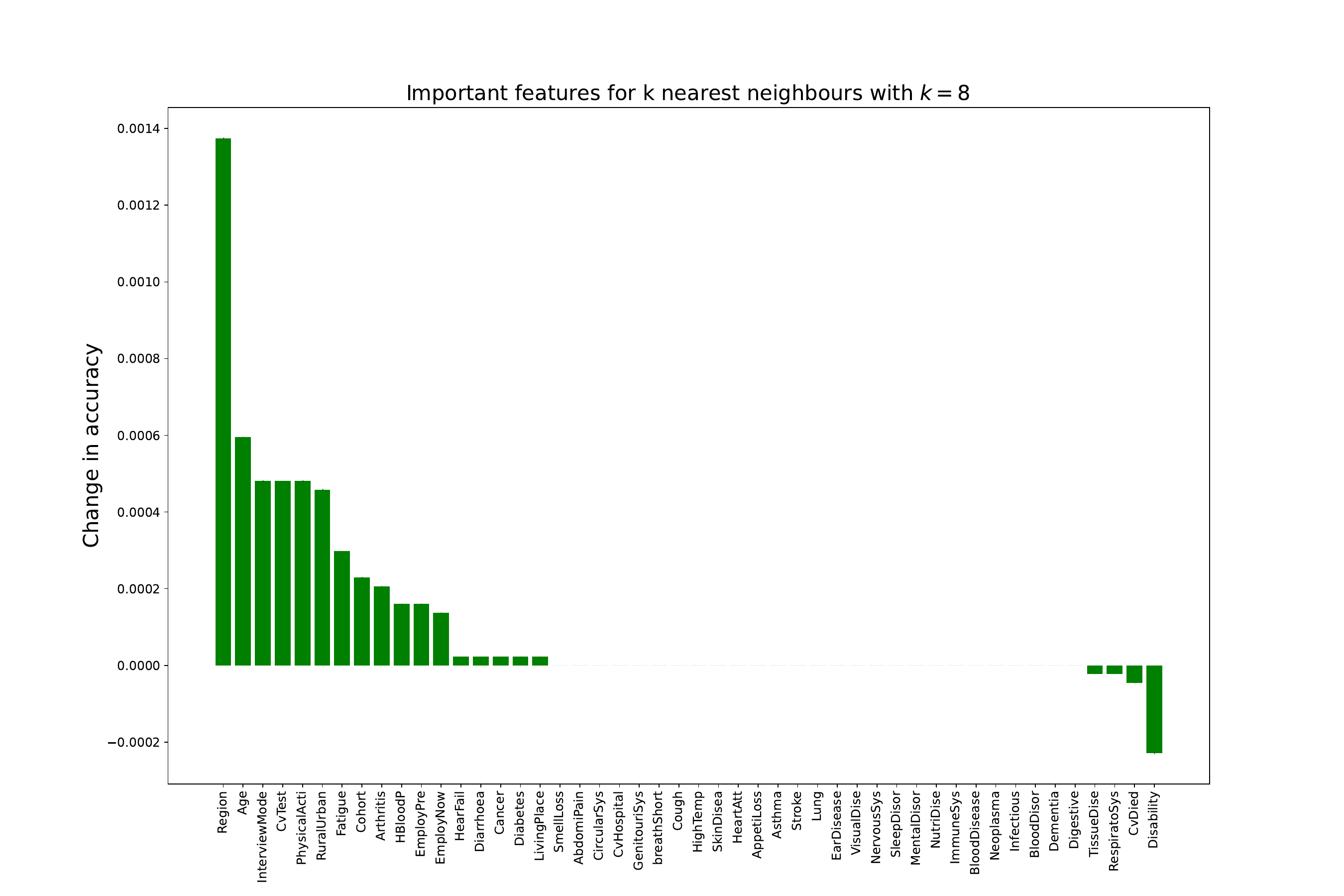}
\includegraphics[scale = 0.3]{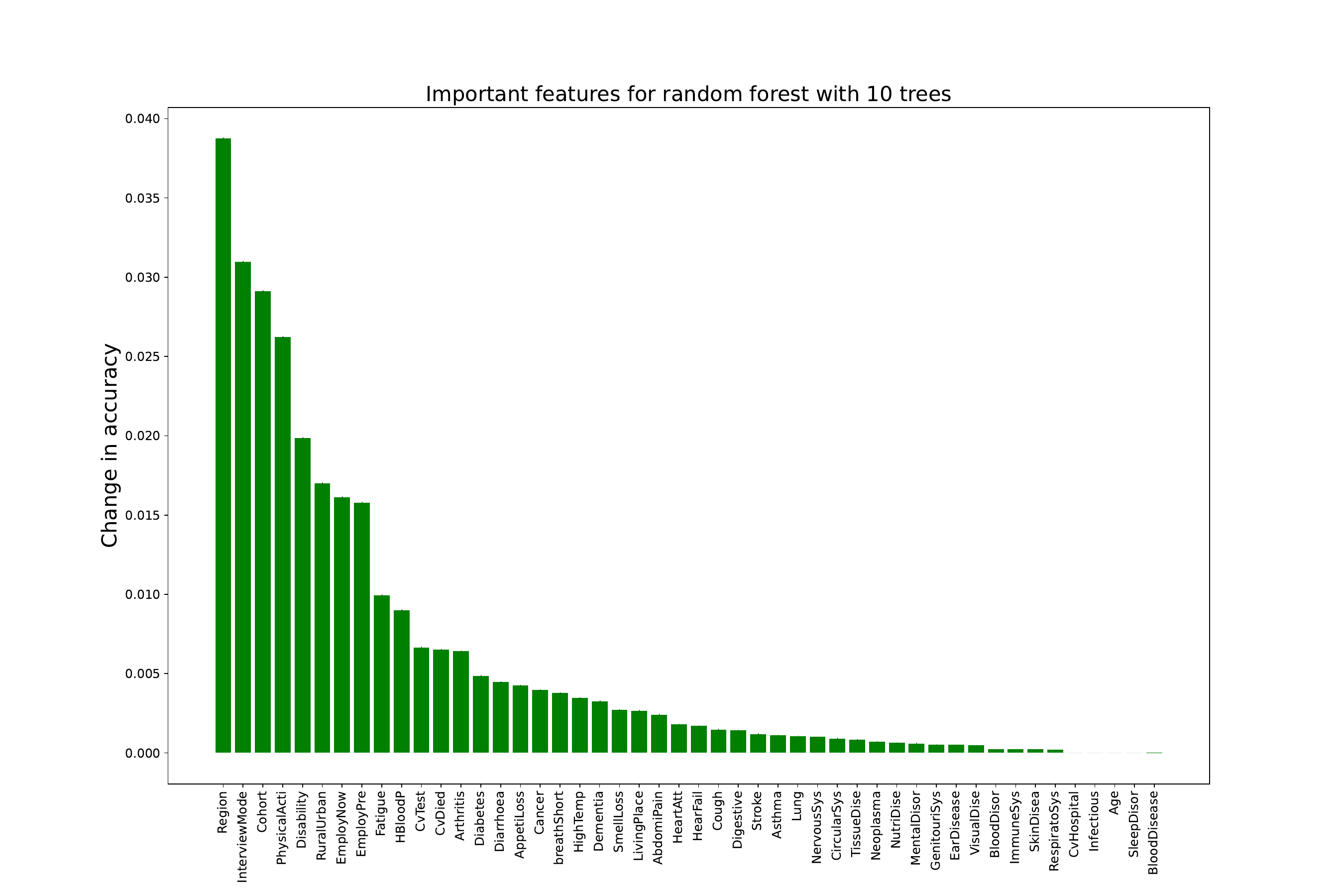}
\caption{Feature importance}
\label{fig:features1}
\end{figure}

\begin{figure}[H]
\centering
\includegraphics[scale = 0.3]{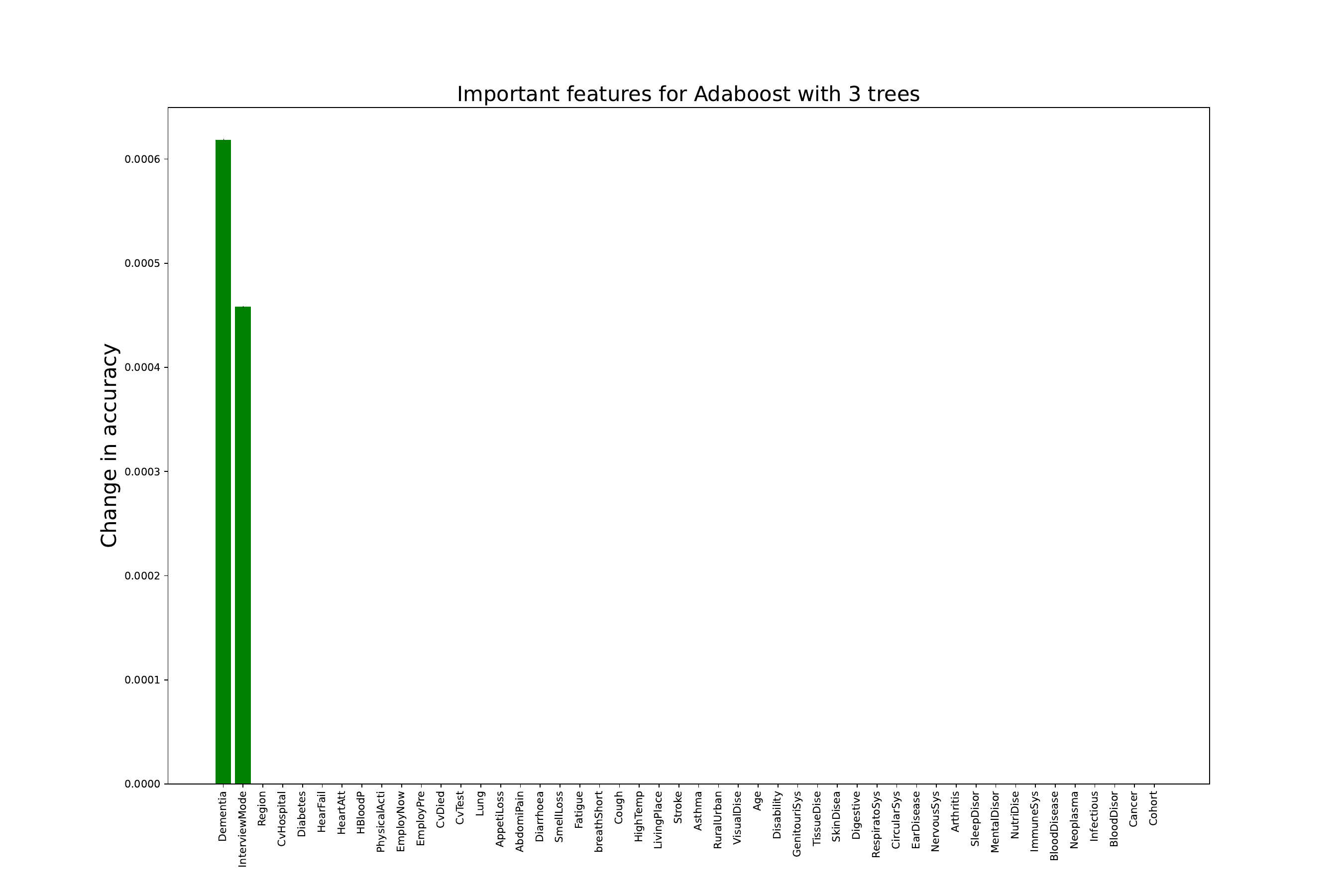}
\includegraphics[scale = 0.3]{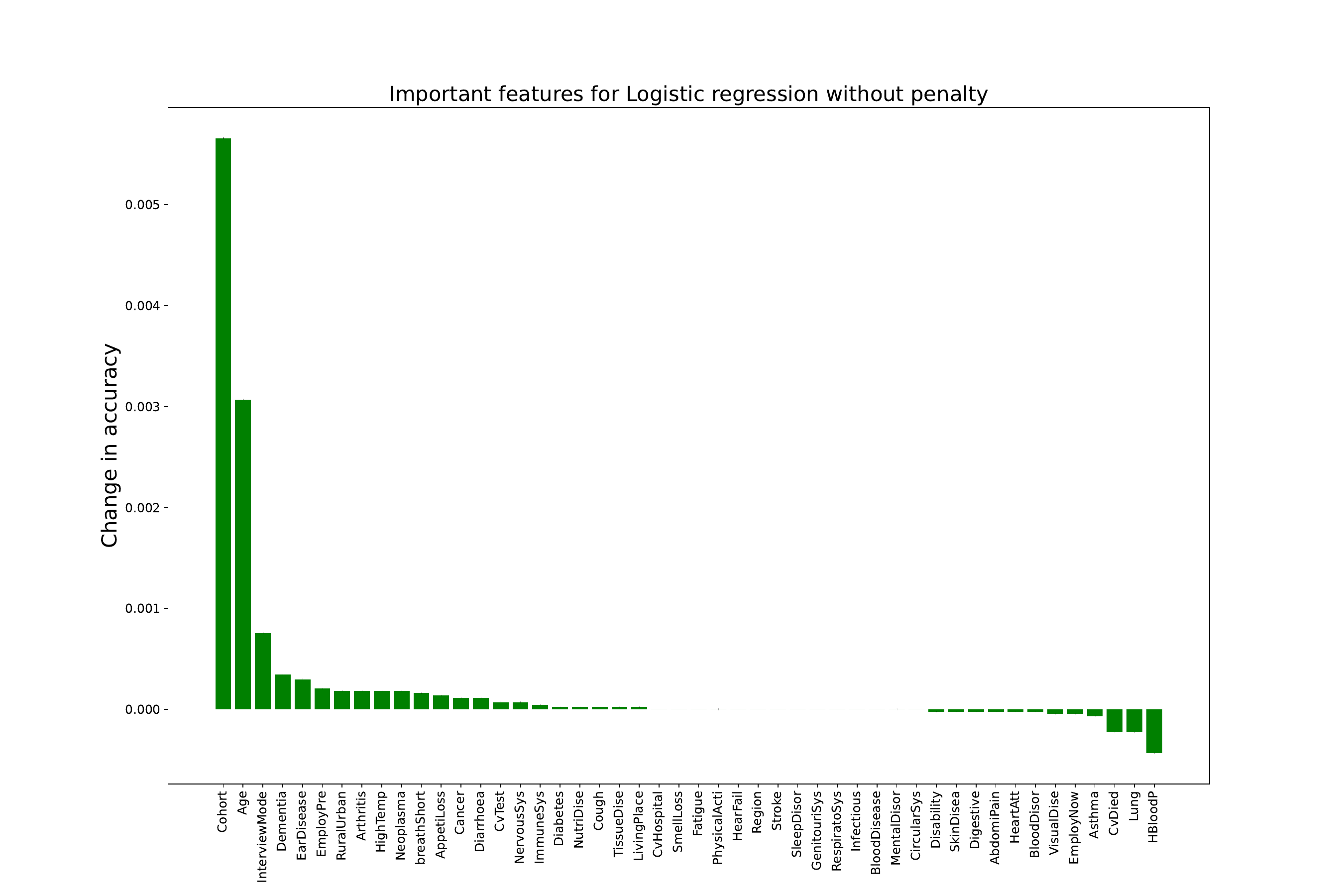}
\caption{Feature importance}
\label{fig:features2}
\end{figure}

\begin{figure}[H]
\centering
\includegraphics[scale = 0.3]{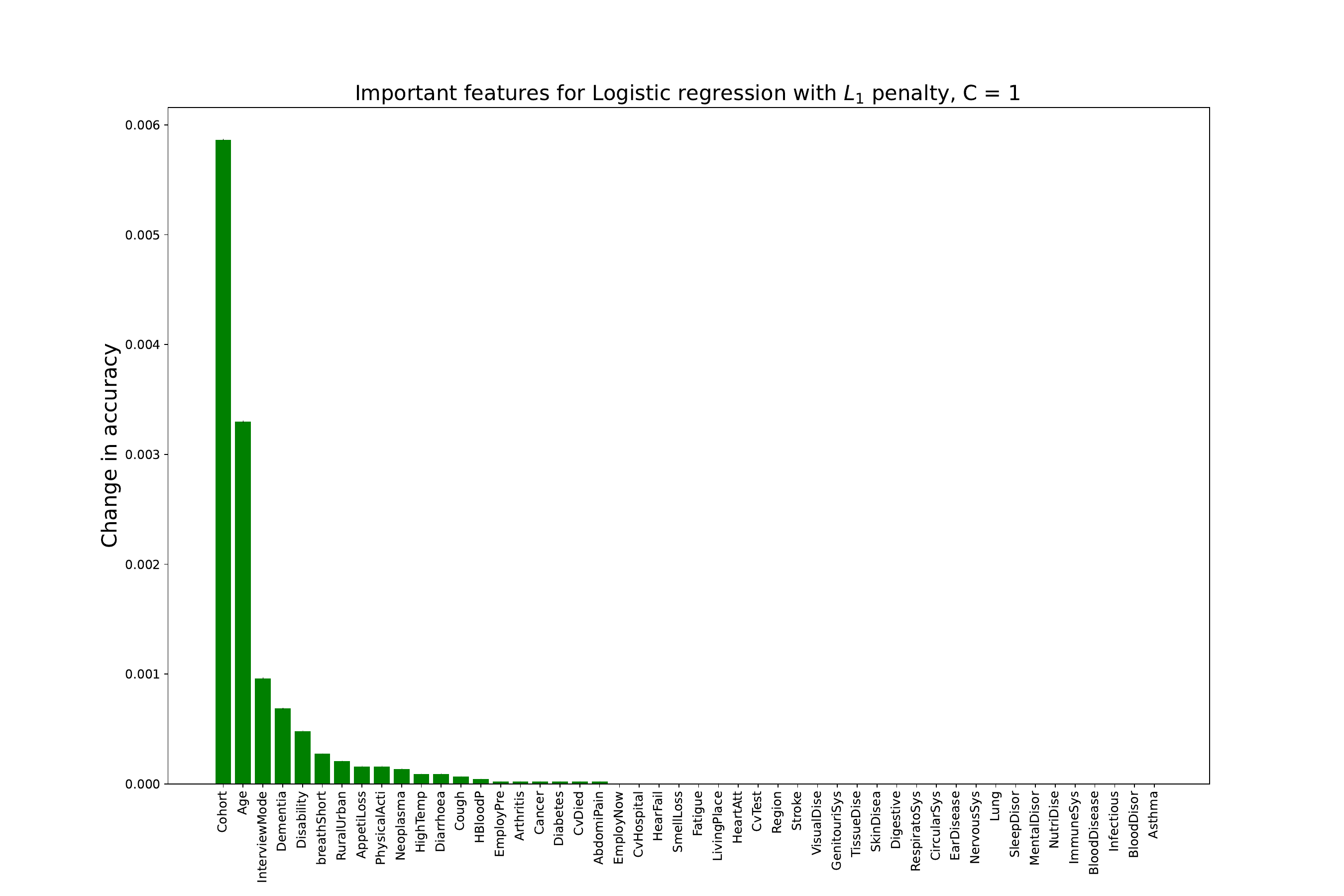}
\includegraphics[scale = 0.3]{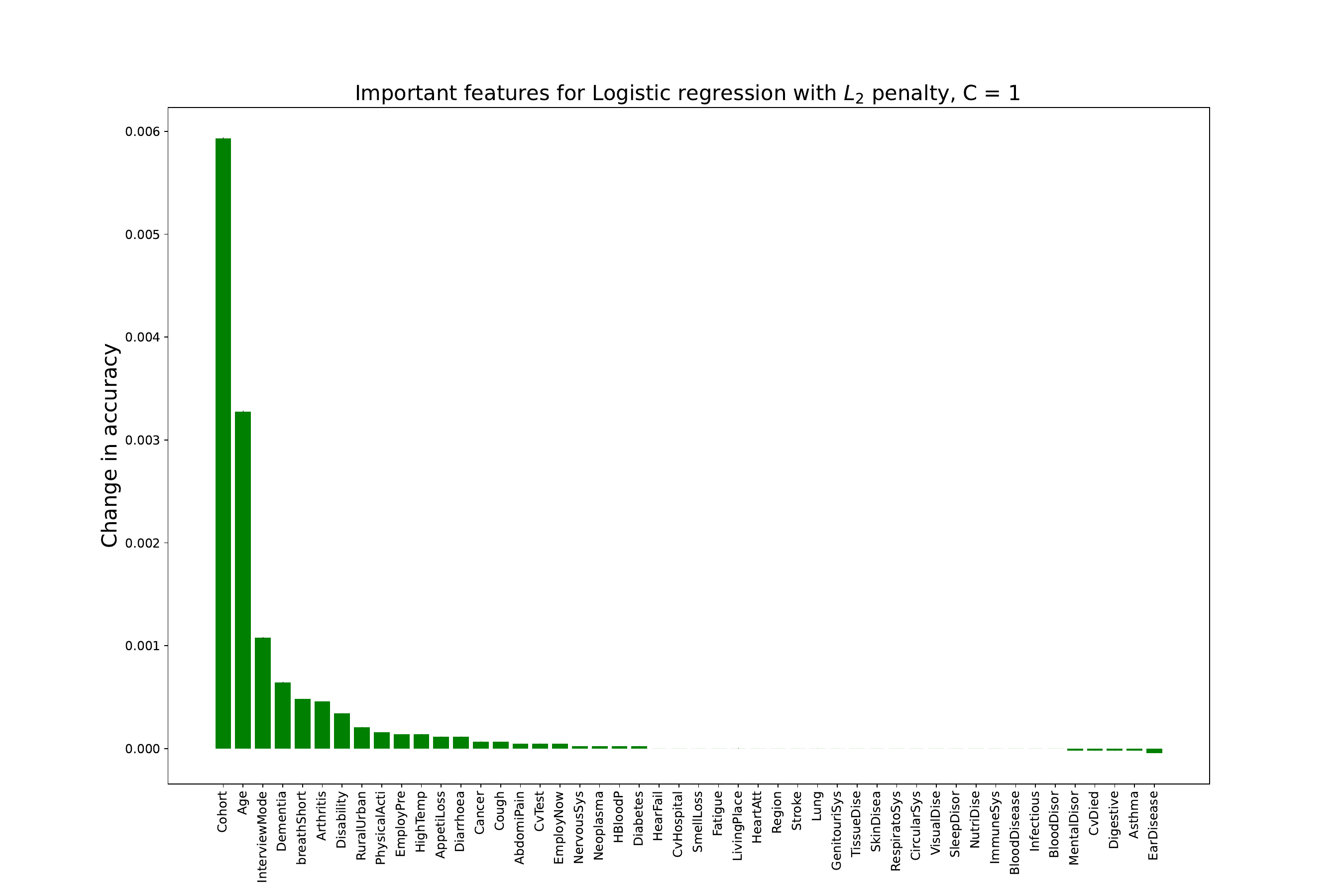}
\caption{Feature importance}
\label{fig:features3}
\end{figure}

\begin{figure}[H]
\centering
\includegraphics[scale = 0.3]{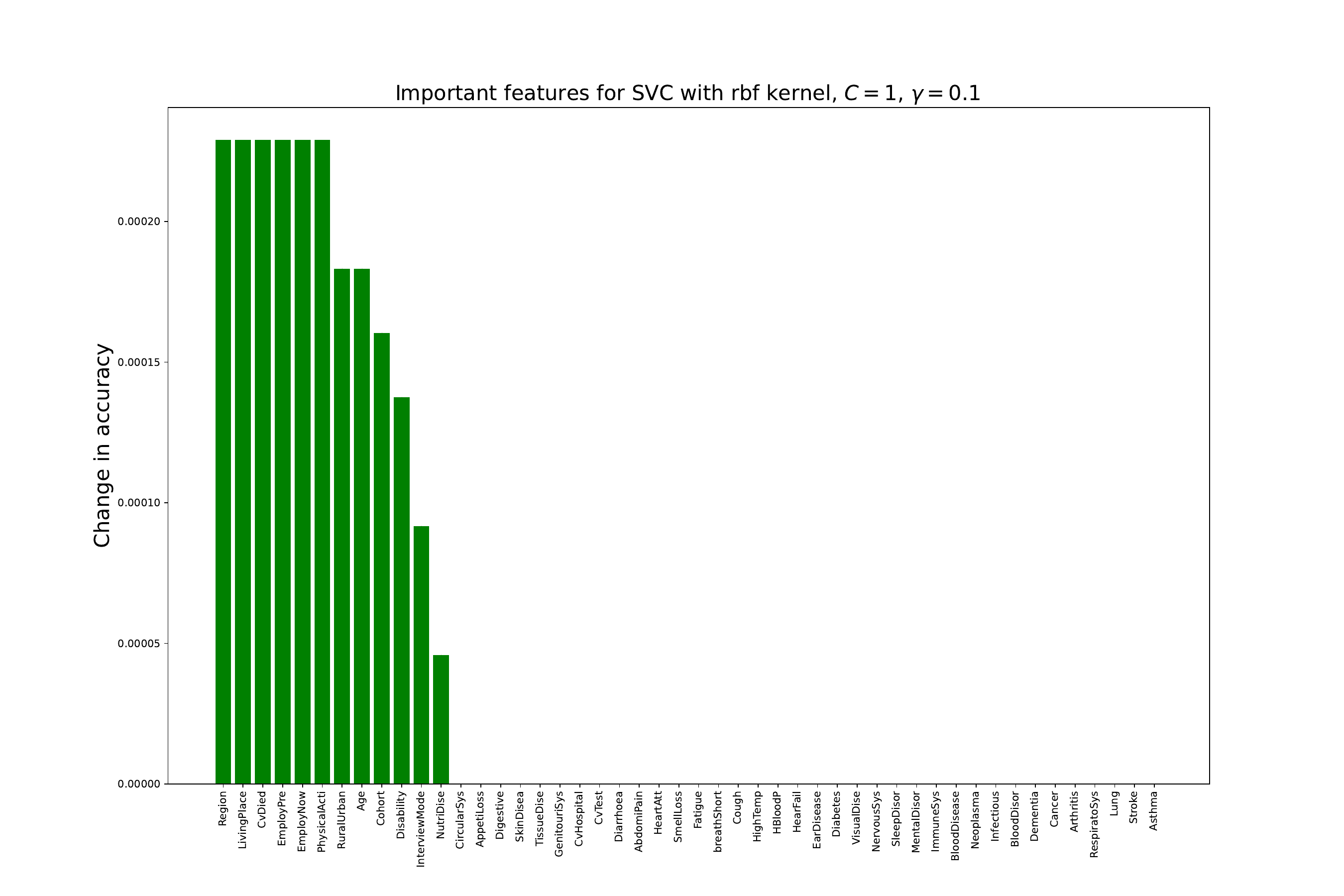}
\includegraphics[scale = 0.3]{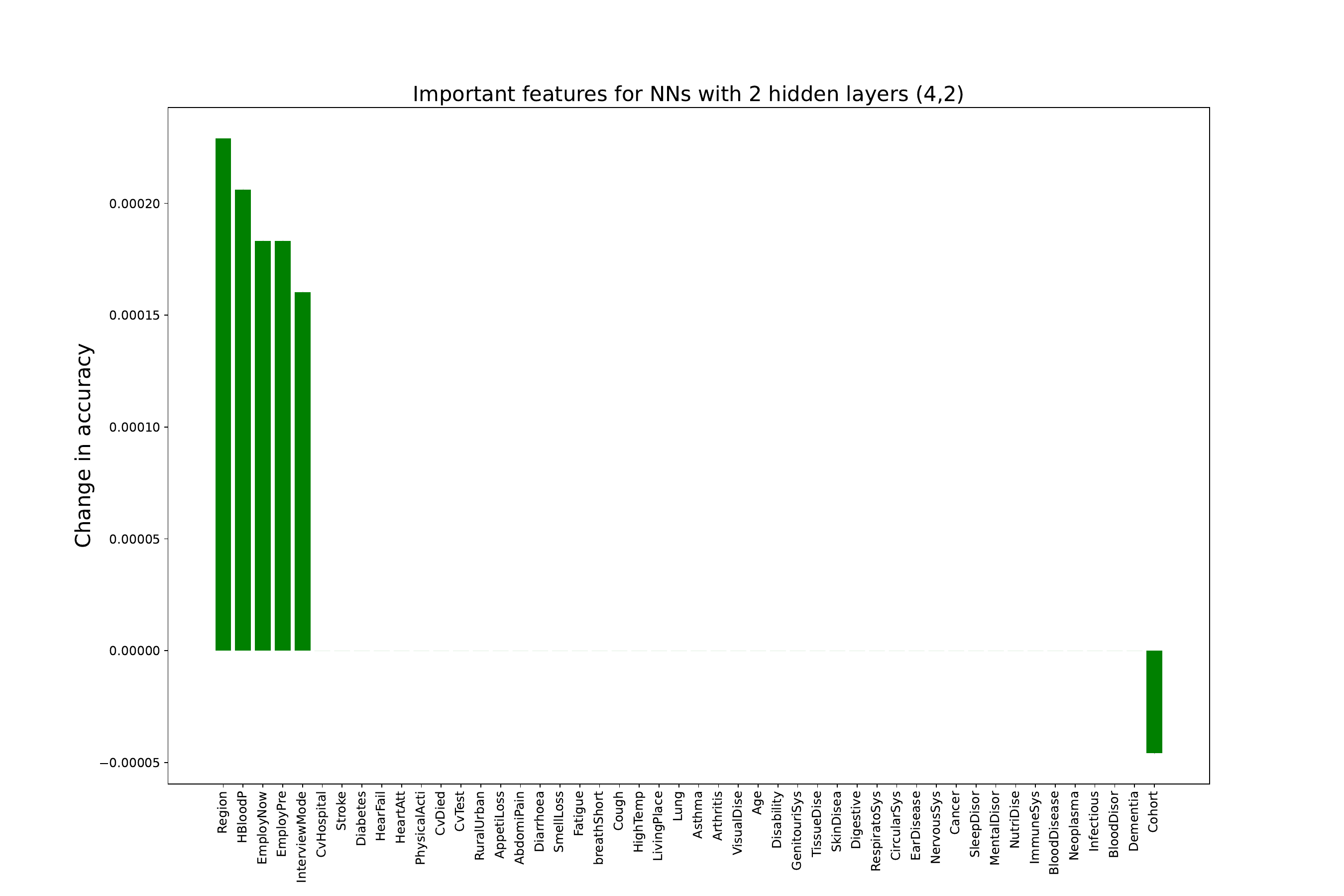}
\caption{Feature importance}
\label{fig:features4}
\end{figure}

\newpage

\section{Appendix}

\subsection{Optimum hypothesis and weight for AdaBoost}
\label{sec:optada}

We start with the loss function in \ref{eq:lossAda}:
\begin{eqnarray}
\label{eq:ada1}
L_t(h) &=& \sum_{i=1}^N \exp\{ - \tilde{y}_i [H_{t-1}(\mathbf{x}_i) + \alpha_t h_t(\mathbf{x}_i)]\}\nonumber\\
&=& \sum_{i=1}^N e^{-\tilde{y}_i H_{t-1}(\mathbf{x}_i)} \, e^{-\tilde{y}_i \alpha_t h_t(\mathbf{x}_i)}.
\end{eqnarray}
We then set $w_{it} = e^{-\tilde{y}_i H_{t-1}(\mathbf{x}_i)} $ and re-write \ref{eq:ada1}, giving
\begin{eqnarray}
L_t(h) = \sum_{i=1}^N w_{it} e^{-\tilde{y}_i \alpha_t h_t(\mathbf{x}_i)}\nonumber.
\end{eqnarray}
We can split this sum as
\begin{eqnarray}
\label{eq:ada2}
L_t(h) = \sum_{i:h(\mathbf{x}_i)=\tilde{y}_i} w_{it} e^{-\alpha_t} + \sum_{i:h(\mathbf{x}_i)\neq \tilde{y}_i} w_{it} e^{\alpha_t}.
\end{eqnarray}
Then the first term in \ref{eq:ada2} can be written as
\begin{eqnarray}
L_t(h) &=& \left( \sum_{i=1}^N w_{it} e^{-\alpha_t} - \sum_{i:h(\mathbf{x}_i)\neq \tilde{y}_i} w_{it} e^{- \alpha_t}\right) + \sum_{i:h(\mathbf{x}_i)\neq \tilde{y}_i} w_{it} e^{\alpha_t}\nonumber\\
&=& \sum_{i=1}^N w_{it} e^{-\alpha_t}  +  \sum_{i:h(\mathbf{x}_i)\neq \tilde{y}_i} w_{it} (e^{\alpha_t} - e^{-\alpha_t}) \nonumber,
\end{eqnarray}
where the first term is independent of $h$ and therefore does not affect optimisation. We then have
\begin{eqnarray}
L_t(h) =  \sum_{i:h(\mathbf{x}_i)\neq \tilde{y}_i} w_{it} (e^{\alpha_t} - e^{-\alpha_t})\nonumber.
\end{eqnarray}
Therefore, given that $e^{\alpha_t} - e^{-\alpha_t}$ is a constant, $L$ is minimised when $h$ is
\begin{eqnarray}
h^{*} = \arg \min \sum_{i=1}^N w_{it} \, I(\tilde{y}_i \neq h(\mathbf{x}_i))\nonumber.
\end{eqnarray}
Next, we find the optimum value of $\alpha$. We start with \ref{eq:ada2} and take a derivative with respect to $\alpha$
\begin{eqnarray}
\frac{\partial}{\partial \alpha_t} L_t(h) &=& \frac{\partial}{\partial \alpha_t} \left(\sum_{i:h(\mathbf{x}_i)=\tilde{y}_i} w_{it} e^{-\alpha_t} + \sum_{i:h(\mathbf{x}_i)\neq \tilde{y}_i} w_{it} e^{\alpha_t}\right)\nonumber\\
&=& - \sum_{i:h(\mathbf{x}_i)=\tilde{y}_i} w_{it} e^{-\alpha_t} +  \sum_{i:h(\mathbf{x}_i)\neq \tilde{y}_i} w_{it} e^{\alpha_t}\nonumber.
\end{eqnarray}
Setting this equal to $0$ and solving for $\alpha_t$, we get
\begin{eqnarray}
\alpha_t^{*} &=& \frac{1}{2} \log \left(\frac{\sum_{i:h(\mathbf{x}_i)= y_i} w_{it}}{\sum_{i:h(\mathbf{x}_i)\neq y_i} w_{it}}\right)\nonumber\\
&=& \frac{1}{2} \log \left(\frac{\sum_{i=1}^N w_{it} - \sum_{i:h(\mathbf{x}_i)\neq y_i} w_{it}}{\sum_{i:h(\mathbf{x}_i)\neq y_i} w_{it}}\right)\nonumber\\
&=&  \frac{1}{2} \log \left(\frac{1 - \frac{\sum_{i:h(\mathbf{x}_i)\neq y_i} w_{it}}{\sum_{i=1}^N w_{it}}}{\frac{\sum_{i:h(\mathbf{x}_i)\neq y_i} w_{it}}{\sum_{i=1}^N w_{it}}}\right)\nonumber\\
&=& \frac{1}{2} \log \left(\frac{1-e_t}{e_t}\right)\nonumber.
\end{eqnarray}

\subsection{Lagrangian and duality}
\label{sec:lag}
Let $L_P$ be the Lagrangian function (primal) and $\lambda$ and $\mu$ be the Lagrangian multipliers (dual variables). We consider the optimisation problem in \ref{eq:svcreg} and write the Lagrangian function as

\begin{eqnarray}
\label{eq:primal}
L_P(\theta, \lambda, \mu) = \frac{1}{2}\|\theta\|^2 + C \sum_{i=1}^N \xi_i - \sum_{i=1}^N \lambda_i [y_i (\theta^T x_i + \theta_0) - (1 - \xi)] - \sum_{i=1}^N \mu_i \xi_i.
\end{eqnarray}
Then, we minimise $L_P$ with respect to $\theta$, $\theta_0$ and $\xi$, giving
\begin{eqnarray}
\label{eq:partial1}
\frac{\partial L_P}{\partial \theta} &=& \theta - \sum_{i=1}^N \lambda_i y_i x_i = 0 \implies \theta = \sum_{i=1}^N \lambda_i y_i x_i\\\label{eq:partial2}
\frac{\partial L_P}{\partial \theta_0} &=&  \sum_{i=1}^N \lambda_i y_i = 0 \\ \label{eq:partial3}
\frac{\partial L_P}{\partial \xi_i} &=& C - \lambda_i - \mu_i = 0 \implies \lambda_i = C - \mu_i, \quad \forall i.
\end{eqnarray}
The dual problem says that the Lagrangian dual objective function $L_D$ is the solution to the minimisation of the primal function, i.e.
\begin{eqnarray}
L_D(\lambda, \mu) = \min_{\theta, \theta_0} L_P(\theta, \lambda, \mu)\nonumber.
\end{eqnarray}
Therefore, substituting \ref{eq:partial1}, \ref{eq:partial2}, and \ref{eq:partial3} in \ref{eq:primal} gives rise to the dual function, giving
\begin{eqnarray}
\label{eq:dual}
L_D(\lambda,\mu) &= &\frac{1}{2} \theta^T\theta + C \sum_{i=1}^N \xi_i - \sum_{i=1}^N \lambda_i y_i \theta^T x_i - \sum_{i=1}^N \lambda_i y_i \theta_0 + \sum_{i=1}^N \lambda_i (1-\xi_i) - \sum_{i=1}^N \mu_i \xi_i\nonumber\\
&=& -\frac{1}{2} \theta^T \theta + \sum_{i=1}^N \lambda_i\nonumber\\
&=& \sum_{i=1}^N \lambda_i - \frac{1}{2} \sum_{i=1}^N \sum_{i=1}^N \lambda_i \lambda_j y_i y_j x_i^T x_j\nonumber\\
&=& \sum_{i=1}^N \lambda_i - \frac{1}{2} \sum_{i=1}^N \sum_{i=1}^N \lambda_i \lambda_j y_i y_j \langle x_i, x_j \rangle.
\end{eqnarray}
Then, we can show that $L_D$ gives a lower bound on the objective function in \ref{eq:svcreg} for any feasible point.  According to the dual problem, we need the best lower bound. Further, according to weak duality theorem, if $\theta^{*}$ is the optimal value of the primal problem and $(\lambda^{*}, \mu^{*})$ are the optimal values of the dual problem, then the optimal value of the dual problem is the best lower bound to the optimal value of the primal problem and hence there exists an optimal value of the dual problem which is equal to an optimal value of the primal problem. Therefore the dual optimisation problem is
\begin{eqnarray}
\max_{\lambda} && \sum_{i=1}^N \lambda_i - \frac{1}{2} \sum_{i=1}^N \sum_{i=1}^N \lambda_i \lambda_j y_i y_j \langle x_i, x_j \rangle\nonumber\\
s.t. && 0 \leq \lambda_i \leq C, \quad i = 1, \dots, N\nonumber\\
&& \sum_{i=1}^N \lambda_i y_i = 0.
\end{eqnarray}  
Solving the dual optimisation problem is simpler than the primal optimisation problem. (See Hastie et al., 2009 and Bierlaire, 2018)

\subsection{Gradient descent and its variants}
\label{sec:gradient}
A common approach to optimisation is taking an initial point and then moving towards the descent direction step by step until it converges to a local minimum. The descent direction can be determined by the gradient or the Hessian. At each iteration the updated point is given by
\begin{eqnarray}
\mathbf{x} := \mathbf{x} + \alpha \mathbf{d}\nonumber
\end{eqnarray}
where $\alpha$ is the learning rate or step size and $\mathbf{d}$ is the descent direction. There are different optimisation methods which differ based on the way $\alpha$ and $\mathbf{d}$ are selected. In gradient descent algorithm, the descent direction is the direction of the steepest descent, i.e. the direction opposite the gradient. Therefore, we have
\begin{eqnarray}
\label{eq:vanila}
\mathbf{x} := \mathbf{x} - \alpha \mathbf{g},
\end{eqnarray}
where $\mathbf{g} = \nabla f(\mathbf{x})$, i.e. the gradient of a function at point $\mathbf{x}$. When the region is relatively flat, gradient descent takes a long time to converge. In that case, we can add a momentum term to \ref{eq:vanila}, giving
\begin{eqnarray}
\mathbf{x} := \mathbf{x} - \alpha \mathbf{g} + \beta \mathbf{v}\nonumber.
\end{eqnarray}
Another variant is when we find the gradient at point $\mathbf{x} + \beta \mathbf{v}$ instead of $\mathbf{x}$. In these two methods, the learning rate is constant. The adaptive subgradient method (adagrad) applies an adaptive learning rate. The updated $\mathbf{x}$ at $k$-th iteration is then given by
\begin{eqnarray}
\mathbf{x}^{k+1}_i = \mathbf{x}^k_i - \eta \mathbf{g}^k_i,
\end{eqnarray}
where $\eta = \frac{\alpha}{\epsilon + \sqrt{\mathbf{s}^k_i}}$, $\mathbf{s}_i = \sum_{j=1}^k (\mathbf{g}_i^{j})^2$, and $\epsilon$ is a small value to prevent division by zero (Duchi et al., 2011). This method is less sensitive to the choice of $\alpha$ and $\alpha$ is usually set to $0.01$. The only problem is that the learning rate is monotonically decreasing and can be very small before convergence. RMSProp is another method that replaces $\mathbf{s}^k_i$ by
\begin{eqnarray}
\mathbf{s}^{k+1} = \gamma \mathbf{s}^k + (1-\gamma) (\mathbf{g}^k \odot \mathbf{g}^k),
\end{eqnarray}
where $\gamma \in (0,1)$ and is usually set to $0.9$ (Tieleman and Hinton, 2012). The adam algorithm discussed in Section \ref{sec:NN} is based on the idea of adagrad and RMSProp. (See also Kochenderfer and Wheeler, 2019 for more details)

\end{document}